\documentclass{jfm}

\usepackage{graphicx}%
\usepackage{natbib}
\usepackage{graphicx}
\usepackage{xargs}                      % Use more than one optional parameter in a new commands
\usepackage[usenames,dvipsnames,svgnames,table]{xcolor}
\usepackage{amsmath}
\usepackage{commath}
\usepackage{amssymb}
\usepackage{amsbsy}
\usepackage{amsmath}
\usepackage{ulem}
\usepackage{hyperref}

\ifCUPmtlplainloaded \else
  \checkfont{eurm10}
  \iffontfound
    \IfFileExists{upmath.sty}
      {\typeout{^^JFound AMS Euler Roman fonts on the system,
                   using the 'upmath' package.^^J}%
       \usepackage{upmath}}
      {\typeout{^^JFound AMS Euler Roman fonts on the system, but you
                   dont seem to have the}%
       \typeout{'upmath' package installed. JFM.cls can take advantage
                 of these fonts,^^Jif you use 'upmath' package.^^J}%
      }
  \else
  \fi
\fi

\ifCUPmtlplainloaded \else
  \checkfont{msam10}
  \iffontfound
    \IfFileExists{amssymb.sty}
      {\typeout{^^JFound AMS Symbol fonts on the system, using the
                'amssymb' package.^^J}%
       \usepackage{amssymb}%
         \let\leq=\leqslant
         \let\geq=\geqslant
      }{}
  \fi
\fi

\ifCUPmtlplainloaded \else
  \IfFileExists{amsbsy.sty}
    {\typeout{^^JFound the 'amsbsy' package on the system, using it.^^J}%
     \usepackage{amsbsy}}
    {\providecommand\boldsymbol[1]{\mbox{\boldmath $##1$}}}
\fi

\newcommand{\vx}{{\mathbf x}}

\begin{document}
% %%%%%%%%%%%%%%%%%%%%%%%%%%%%%%%%%%%%%%%
\title{Dispersion of motile bacteria in\\ a porous medium}
%%%%%%%%%%%%%%%%%%%%%%%%%%%%%%%%%%%%%%%%
 
\author{Marco Dentz\aff{1}
  \corresp{\email{marco.dentz@csic.es}},
  Adama Creppy\aff{2}, Carine Douarche\aff{2}, Eric Cl\'ement\aff{3,4} \and Harold Auradou\aff{2}}

\affiliation{\aff{1}Spanish National Research Council (IDAEA-CSIC), Barcelona, Spain
\aff{2} Université Paris-Saclay, CNRS, FAST, 91405, Orsay, France \aff{3}Laboratoire PMMH-ESPCI Paris, PSL Research University, Sorbonne University, University Paris-Diderot, 7, Quai Saint-Bernard, Paris, France.
\aff{4} Institut Universitaire de France (IUF).}
\date{\today}

 \maketitle

\begin{abstract}
{Understanding flow and transport of bacteria in porous media is crucial to
  technologies such as bioremediation, biomineralization or enhanced oil
  recovery. While physicochemical bacteria filtration is well-documented, recent
  studies showed that bacterial motility plays a key role in the transport
  process. Flow and transport experiments performed in microfluidic chips
  containing randomly placed obstacles confirmed that
  the distributions of non-motile bacteria stays compact, whereas for the
  motile strains, the distributions are characterized by both significant
  retention as well as fast downstream motion. For motile bacteria, the detailed
  microscopic study of individual bacteria trajectories reveals two salient
  features: (i) the emergence of an active retention process triggered by
  motility, (ii) enhancement of dispersion due to the exchange between fast flow
  channels and low flow regions in the vicinity of the solid grains. We propose
  a physical model based on a continuous time random walk approach. This
  approach accounts for bacteria dispersion via variable pore-scale flow
  velocities through a Markov model for equidistant particle speeds. Motility of
  bacteria is modeled by a two-rate trapping process that accounts for the
  motion towards and active trapping at the obstacles. This approach captures
  the forward tails observed for the distribution of bacteria displacements, and
  quantifies an enhanced hydrodynamic dispersion effect that originates in the
  combined effect of pore-scale flow variability and bacterial motility. The model
  reproduces the experimental observations, and predicts bacteria dispersion and
  transport at the macroscale. }
\end{abstract}

\vspace{\baselineskip}

%%%%%%%%%%%%%%%%%%%%%%%%%%%%%%%%%%%%%%%%%
\section{Introduction}
%%%%%%%%%%%%%%%%%%%%%%%%%%%%%%%%%%%%%%%%%
Bacteria are the cause of many diseases and some of them, such as cholera, are spread by contaminated water. In the 19th century, this problem led
to the development of drinking water systems separated from wastewater
and motivated Darcy to formulate the basic equations describing the flow
of a fluid in a porous medium~\cite[][]{darcy1856fontaines}. Since then, bacteria transport and filtration through porous media has
remained a field of intense research. However, still many practical challenges are concerned with difficulties for macroscopic standard models to provide a reliable and quantitative picture of the dispersion of bacteria transported by flow in porous media. For instance, \citet{Hornberg1992} published a study comparing the bacterial effluent curves with those of a classical filtration model including fluid convection and absorption-desorption kinetics. The model allows for a good adjustment of the long time tail of the bacteria concentration curves whereas the model gives disappointing predictions for the breakthrough curves at short times. Subsequent studies have sought to identify the influence of flow or physico-chemical conditions on the model parameters. Although little consideration was given to bacterial motility, it came out that this parameter could be crucial to better understand dispersion and retention processes~\cite[][]{McCaulou1994,Hendry1999,Camesano1998,Guangming2005,Walker2005,Liu2011,Stumpp2011,Zhang2021}. Recent studies support the idea that the swimming capacity of the bacteria allows them to explore more of the porosity~\cite[][]{Becker2003,Liu2011}. For instance, by performing flow experiments with motile and non-motile bacteria in a fracture, \cite{Becker2003} recovered at the outlet about $3 \%$ of the non-motile bacteria and only $0.6\%$ of similar but motile bacteria. The mass loss of motile bacteria was explained by the fact that motility eases the diffusion into stagnant fluid resulting in a greater residence time in the porosity and close to grain surfaces. As a consequence motile bacteria are more likely to be filtered. This conclusion seems however
inconsistent and in contradiction with earlier observations \cite{Hornberg1992} and ~\cite{Camesano1998} reporting less adhesion to soil grains at low fluid velocity.

Microfluidic technology offers a unique experimental method to directly visualize the behavior of bacteria inside pores. Even when using simple geometries such as channels with rectangular cross sections
researchers observed non trivial behavior of bacteria in a flow like upstream motions~\cite[][]{Kaya2012}, back-flow low along corners~\cite[][]{Figueroa2015} eventually leading to large scale "super-contamination" \cite[][]{Figueroa-Morales2020}, 
transverse motions due to chirality-induced rheotaxis ~\cite[][]{Marcos2012,Jing2020}
% G.Jing, A. Zöttl, E. Clément, A. Lindner, Chirality-induced bacterial rheotaxis in bulk shear flows, Science Advances, 6, eabb2012 (2020).
and oscillations along the surfaces~\cite[][]{Mathijssen2019}. Those observations %in microfluidic channels
revealed that the dependence of the bacteria orientations on fluid
shear adds new elements that further complicate the transport description. Some studies also point out that this dependence might affect the macroscopic transport of motile bacteria suspensions. This was revealed by the experimental study of \cite{Rusconi2014}. In this work, the bacterial concentration profile across the width of a microfluidic channels was recorded as function of flow velocity. When flow was increased and concomitantly the shear rate, they observed a depletion of the central part of the profile that they attributed to a transverse flux of bacteria from low shear to high shear regions located near the surfaces~\cite[][]{Rusconi2014}. Motility was also observed to lead to bacteria accumulation at the rear of a constriction \cite[][]{Altshuler2013} or downstream circular obstacles~\cite[][]{Mino2018,Secchi2020,Lee2021}. Addition
of pillars to microfluidic rectangular channels offers the possibility to design model bi-dimensional heterogeneous porous system suited to explore the influence of flow heterogenities and pore structures on the transport and retention of bacteria~\cite[][]{Creppy2019,Dehkharghani2019,Scheidweiler2020,Secchi2020,deAnna2020}. This approach allows for tracking of individual bacteria trajectories and the measurement of statistical quantities leading to significant progresses towards the understanding and modeling of bacteria transport and dispersion at a macroscopic scale. They all point out that motility has two major impacts, it increases the residence time close to the grains and in regions of low velocity and favors the adhesion~\cite[][]{Scheidweiler2020}. The increase of probability to be close to the grains was recently observed in periodic porous media~\cite[][]{Dehkharghani2019}. The effect on  the macroscopic longitudinal dispersion was then investigated numerically using Langevin simulations. Their study revealed a strong enhancement of the dispersion coefficient particularly when the flow is aligned along the crystallographic axis of the porous medium. In this case, the dispersion coefficient is found to increase like the flow velocity to the power 4 instead of a power 2 as classically obtained for Taylor dispersion. Those examples also show that an accurate macroscopic transport model based on the pore scale observations suited to predict the fate of motile bacteria transported in a porous flow is still missing.

% \citet{Scheidweiler2020} also point out that the continuous time
% random walk (CTRW) approach is able to model successfully transport and filtration processes of bacteria transferred 
% in  microfluidic chips, allowing then to determine quantitatively the
% absorption rates of motile and of non-motile
% bacteria~\cite{Scheidweiler2020}. 
 
 Current approaches to quantify the impact of motility on bacteria dispersion use the generalized Taylor dispersion approach developed by~\cite[][]{brenner}, which is based on volume averaging of the pore-scale Fokker-Planck equation that describes the distribution of bacteria position and orientation~\cite[][]{Alonso2019}. This approach lumps the combined effect of pore-scale flow variability and motility into an asymptotic hydrodynamic dispersion coefficient. Therefore, it has the same limitations as macrodispersion theory in that it is not able to account for non-Fickian transport features such as forward tails in the distribution of bacteria displacements and non-linear evolution of the displacement variance. The data-driven approach of~\citet{Massoudieh2018} mimics the run and tumble motion of the bacteria by a mesoscopic stochastic model that represents the motile velocity as a Markov process characterized by an empirical transition matrix, but does not provide an upscaled model equation for bacteria dispersion.   
 
 In this paper, our aim is to develop a physics-based mesoscale model for bacteria motion, and derive the upscaled transport equations, by explicitly representing pore-scale flow variability and motility, and their combined impact on bacteria dispersion. In order to understand and quantify the role of motility, we used the experimental data obtained by \citet{Creppy2019}. Because these experiments were performed at various flow rates and with motile and non-motile bacteria, this data set offers the possibility to investigate the effect of the flow velocity on bacterial motion. We use a continuous time random walk (CTRW) approach~\cite[][]{morales2017,Dentz2018jfm} to model the advective displacements of bacteria along streamlines at variable flow velocities, while the impact of motility is represented as a two-rate trapping process. A similar travel time based approach was used by~\citet{deJong1958} and~\citet{saffman1959} to quantify hydrodynamic dispersion coefficients in porous media.
 
 %In our  approach, the characteristics of the flow field, i.e, the velocity
 %distribution, correlation length of the velocity field and the
 %tortuosity are first determined from experiments performed with 
 %non-motile bacteria by adjusting a Continuous Time Random Walk (CTRW)
 %model. Motility is modeled by a two-rate trapping process that is characterized by 
 %trapping and release rate that accounts for motion towards and dwelling 
 %at the grains. 
 
 %A fairly good fit of the experimental data is achieved for
 %untrapping and trapping rates proportional to the
 %local flow velocity with a prefactor that depends on the ratio
 %between the flow velocity and the swimming velocity of the bacteria.
 %The prefactor accounts for a more efficient trapping when the flow
 %velocity is of the order of the swimming velocity.
 
 The paper is organized as follows. Section~\ref{sec:exp} reports on the experimental data for the displacement and velocity statistics of motile and non-motile bacteria.  Section~\ref{sec:nm} analyzes transport of non-motile bacteria, which can be considered as passive particles. Thus, we use a CTRW approach, which is suited to quantify the impact of hydrodynamic variability on dispersion. This approach forms the basis for the derivation of a CTRW-based model for the transport of motile bacteria in Section~\ref{sec:m}, which accounts for both hydrodynamic transport and motility. {A central element here is to consider and quantify the motility based motion of bacteria toward the solid as an effective trapping mechanism.}
%%%%%%%%%%%%%%%%%%%%%%%%%%%%%%%%%%%%%%%%%%%%%%%%%%%
\section{Experimental data\label{sec:exp}}

%%%%%%%%%%%%%%%%%%%%%%%%%%%%%%%%%%
\begin{figure}
  \centering
  \includegraphics[width=\textwidth]{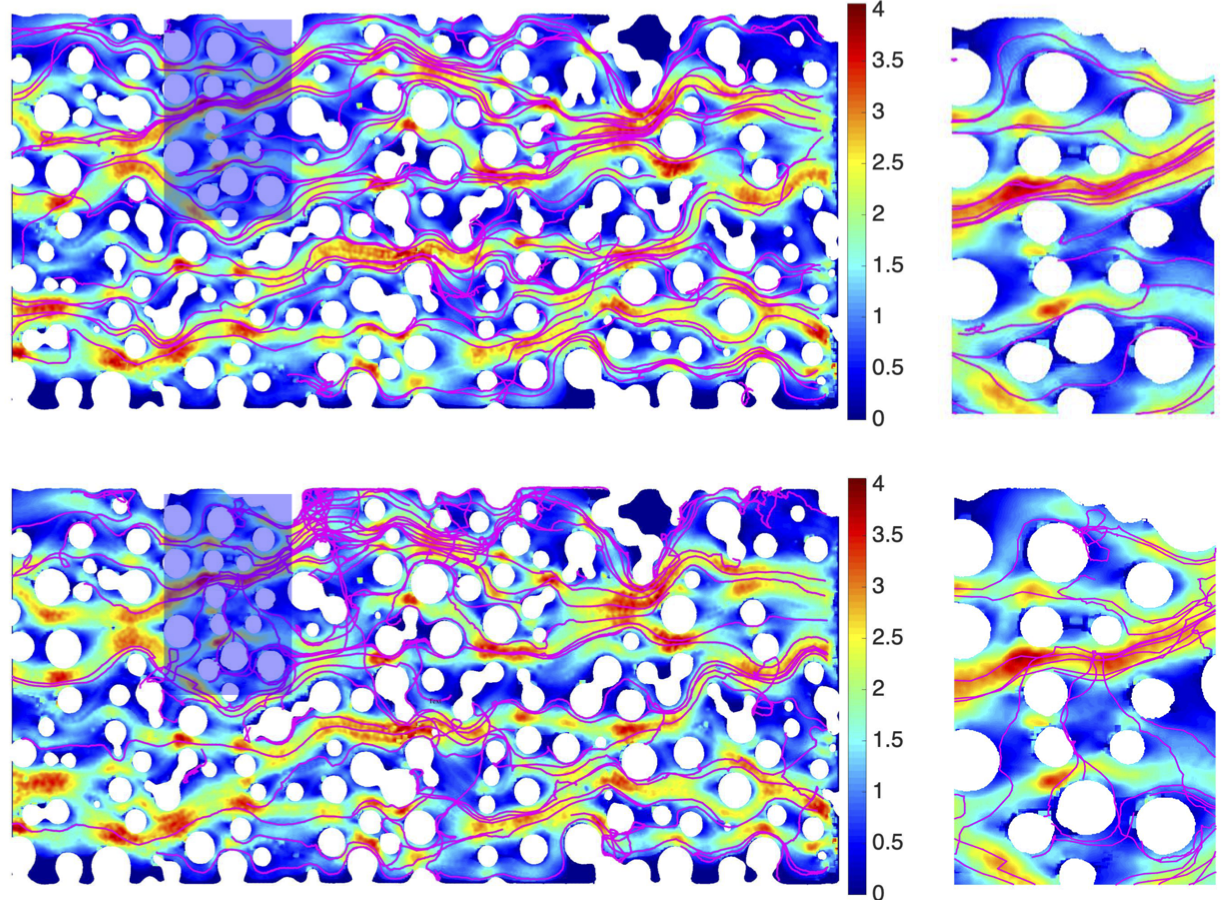}
  \caption{Trajectories of motile bacteria at mean velocities (top) $u_m = 98\mu$m/s
 and (bottom) $u_m = 43 \mu$m/s. The shaded area indicates the zoom in the right panel. The color code corresponds to the local fluid velocities with respect to the mean. Velocity data were obtained by tracking passive particles in the flow. }
  \label{fig:experiment}
\end{figure}
%%%%%%%%%%%%%%%%%%%%%%%%%%%%%%%%%% 

We use the extensive data set of~\citet{Creppy2019} for the displacements of
non-motile and motile bacteria in a model porous medium consisting of vertical
cylindrical pillars placed randomly in a Hele-Shaw cell of height $h = 100 \mu$m,
also termed the grains in the following. The pillar diameters were chosen
randomly from a discrete distribution (20, 30, 40 and 50
$\mu$m) with mean $\ell_0 = 35 \mu$m, which is about $1/3$ of the cell height.
The grains filled the space with a volume fraction of $33 \%$. This idealized
model porous medium shares some characteristics with natural media in channel
height and grain size~\cite[][]{Bear:1972}. A fluorescent Escherichia coli RP437
strain is used to facilitate optical tracking. Details on the
microfluidic experiments are given in \citet{Creppy2019}. The raw trajectory data were
reanalyzed for this study. We consider
data from $7$ experiments that are characterized by the mean streamwise
velocities of the non-motile bacteria, which are $u_m = 18, 43, 66, 98, 113,
139$ and $197\mu$m/s. In each experiment the motion of both motile and
non-motile bacteria are considered. In the following, we refer to the
experiments as $18$, $43$, $66$ etc. according to the respective mean velocity.
We choose the average grain diameter and the
average absolute value of the particle velocity along the flow direction $u_m$
to define the characteristic advection time $\tau_v = \ell_0/u_m$.
%%%%%%%%%%%%%%%%%%%%%%%%%%%%%%%%%%%%%%%%%%%%%%
\subsection{Displacement moments and propagators}
%%%%%%%%%%%%%%%%%%%%%%%%%%%%%%%%%%%%%%%%%%%%%%
Particle trajectories $\vx(t) = [x(t),y(t)]$ of different lengths and
duration are recorded, along which velocities are sampled, and from which the displacement moments and propagators are determined. Figure~\ref{fig:experiment} illustrates trajectories of non-motile and motile bacteria from the microfluidic experiments. We
focus on displacements along the mean flow direction, which is aligned with the $x$-direction of the coordinate system.
Particle displacements are calculated by
\begin{align}
\Delta x(t_n) = x(t_0 + t_n) - x(t_0),
\end{align}
where $x(t_0)$ is the starting position of the trajectory at time
$t_0$ and $t_n = n \Delta t$ are subsequent sampling times. The time
increment $\Delta t$ is given by the inverse framerate of the camera.  
The displacement moments are determined by averaging over all particle
trajectories
\begin{align}
m_j(t_n) = \frac{1}{N_t} \sum\limits_{k=1}^{N_t} \Delta x_k(t_n)^{j},
\end{align}
where $N_t$ denotes the number of tracks, and subscript $k$ denotes the $k$th trajectory. The displacement variance is defined in terms of the first and second displacement moments by
\begin{align}
    \sigma^2(t_n) = m_2(t_n) - m_1(t_n)^2.
\end{align}
The propagators or displacement distribution is defined by 
\begin{align}
p(x,t_n) =  \frac{1}{N_t} \sum\limits_{k=1}^{N_t} \frac{\mathbb
  I\left[x < \Delta x_k(t_n) \leq x + \Delta x \right]}{\Delta x},
\end{align}
where $\mathbb I(\cdot)$ is the indicator function, which is $1$ if
the argument is true and $0$ else, $\Delta x$ is the size of the
sampling bin. Note that the number of tracks decreases with track
length and sampling time $t_n$, see the discussion in Appendix~\ref{app:tracks}.  
%%%%%%%%%%%%%%%%%%%%%%%%%%%%%%%%%%%%%%%%%%%%%%
\subsection{Velocity statistics}
%%%%%%%%%%%%%%%%%%%%%%%%%%%%%%%%%%%%%%%%%%%%%%

%%%%%%%%%%%%%%%%%%%%%%%%%%%%%%%%%%%%%%%%%%%
\begin{figure}
  \centering
\includegraphics[width=0.45\textwidth]{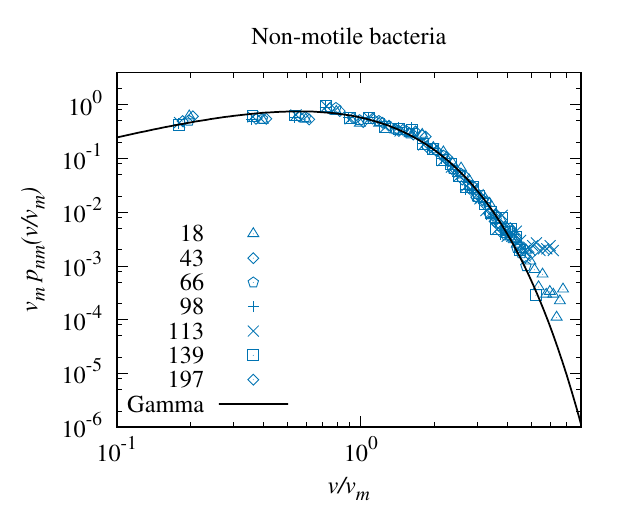}
\includegraphics[width=0.45\textwidth]{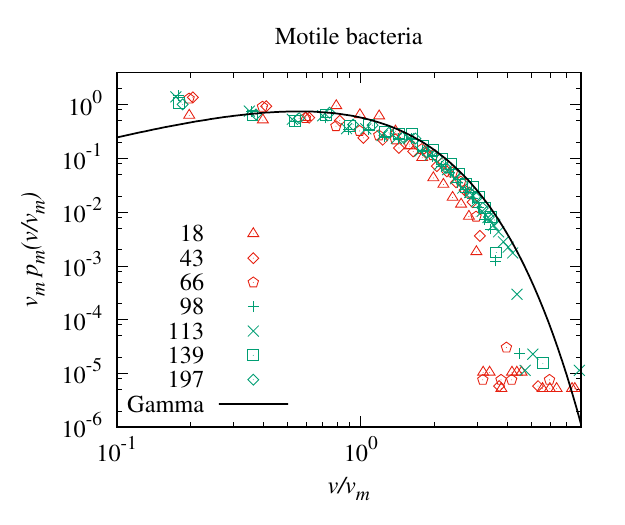}
  \caption{Speed distributions for (left) non-motile and (right) motile bacteria
    for different flow rates rescaled by the mean $v_m$ of the respective
    non-motile speed distributions. The solid black lines in both figures
    denotes the analytical approximation by the Gamma distribution~\eqref{eq:gamma} of the speed
    distribution for the non-motile bacteria for $\alpha = 2.25$. The legend indicates the
    experiments, which are identified by the mean streamwise velocity of
    non-motile bacteria in $\mu$m/s.}
 \label{fig:vpdf}
\end{figure}
%%%%%%%%%%%%%%%%%%%%%%%%%%%%%%%%%%%%%%%%%%%

Particle velocities $\mathbf u(t) = [u_x(t),u_y(t)]$ are obtained from the particle displacements between
subsequent images,
\begin{align}
u_x(t) = \frac{x(t+\Delta t) - x(t)}{\Delta t}, && u_y(t) = \frac{y(t+\Delta t)
  - y(t)}{\Delta t}. 
\end{align}
The particle speed is defined by $v(t) = \sqrt{u_x(t)^2 +
  u_y(t)^2}$. The mean particle velocity in the following is denoted by
$\langle {\mathbf u}(t) \rangle = (u_m,0)$. The mean speed is
denoted by $\langle v(t) \rangle = v_m$. Averages are taken over all
tracks and sampling times. The speed PDFs are obtained by sampling over all
trajectories and sampling times.

Figure~\ref{fig:vpdf} shows the probability
density functions (PDFs) of particle speeds for the non-motile and motile
bacteria, denoted by $p_{nm}(v)$ and $p_m(v)$, respectively, rescaled by the
mean speed $v_m$ of the non-motile bacteria. Non-motile bacteria can be
considered passive tracer particles. Thus, the speed distributions of non-motile
bacteria serves as a proxy for the Eulerian flow speed distribution, that is,
$p_{nm}(v) \equiv p_e(v)$ which is supported by the fact that the rescaled data
collapse on the same curve. The
non-dimensional speed data is well represented by the  Gamma distribution
\begin{align}
\label{eq:gamma}
p_e(v) = \left(\frac{v \alpha}{v_m}\right)^{\alpha-1} \frac{\alpha \exp(-v\alpha/v_m)}{v_m \Gamma(\alpha)}, 
\end{align}
for $\alpha = 2.25$. Speed distributions in
porous media are often characterized by exponential or stretched
exponential decay for $v > v_m$ and power-law behaviors at low flow
speeds. Similar speed distributions have been reported in experimental
particle tracking data~\cite[][]{holzner2015intermittent, morales2017,
  alim2017, carrel2018, souzy2020} and from numerical simulations of
pore-scale flow~\cite[][]{siena2014relationship,Matyka2016,deannaprf2017,Aramideh2018,Dentz2018jfm}.

The right panel of Figure~\ref{fig:vpdf} shows the speed PDFs for the motile
bacteria rescaled by the mean speed of the respective non-motile bacteria,
together with the Gamma distribution given in Eq.~\eqref{eq:gamma}, which models
the non-motile speed PDFs. The global shapes of the rescaled speed PDFs
for the motile bacteria are very similar to the speed PDF for the non-motile
bacteria represented by the Gamma distribution. However, they are shifted towards smaller
values when compared to the non-motile bacteria, with a small peak at low values,
which can be related to bacteria motion along the grains. The
speed PDFs with $u_m \geq 98 \mu$m/s scale with the mean speed $v_m$ and
group together above all at intermediate and small speeds. The speed
PDF of the motile bacteria measures the combined speed of
the flow field and bacteria motility. The fact that the speed PDFs
collapse when rescaled by the respective mean flow speeds indicates
that bacteria motion scales with the flow speed. This seems to be
different for the speed PDFs for $u_m \leq 66 \mu$m/s. The
PDFs are more scattered and shifted towards smaller values compared to
the speed PDFs for the high flow rates. 

Particle trajectories are tortuous due to pore and velocity structures,
and thus are longer than the corresponding linear distance. The ratio
between the average trajectory length of the non-motile bacteria and
the linear length in mean flow direction defines the tortuosity
$\chi$. It can be quantified by the ratio between the mean flow speed
$v_m$ and the mean flow velocity $u_m$ as~\cite[][]{kop1996,
  Ghanbarian2013, puyguiraud2019upscaling}
\begin{align}
\chi = \frac{v_m}{u_m},
\end{align}
We obtain from the velocity data at all flow rates, tortuosity values between
$\chi = 1.17$ and $1.23$. 
%In the following, distances are measured in
%units of the pore length $\ell_0$, times are given in units of advection time $\tau_v =
%\ell_0/u_m$ and velocities are normalized by the mean flow velocity $u_m$.

%%%%%%%%%%%%%%%%%%%%%%%%%%%%%%%%%%%%%%%%%%%%%%%%%%%%%%%%%%%
\begin{table}
%\vspace{-1.5em}
\begin{center}
     \begin{tabular}{r p{10cm} }
%\toprule
$\ell_0$ & grain size\\
$\ell_c$ & characteristic persistence length of particle speeds\\
$\ell'_c$ & \textcolor{black}{coarse-graining length} \\
$v_0$ & magnitude of the swimming velocity of the bacteria\\  
$\mathbf u$ & velocity of non-motile bacteria\\
%$v$ & magnitude of the particle velocity\\ 
$v$ & $=|\mathbf u|$, speed of non-motile bacteria\\
$v_m$ & $=\langle v \rangle$, average speed\\
$u_m$ & $=\langle u_x \rangle$, average streamwise velocity\\
%average absolute value of the x component of the particle velocity. $u_m$ is sometimes referred to the average linear velocity, seepage velocity or average interstitial velocity.\\
$\tau_v$ & $=\ell_0/u_m$, advection time\\
$\chi$ & $=v_m/u_m$, tortuosity\\
$\tau_c$ & characteristic trapping time\\
$\gamma$ & trapping rate\\
$D_{nm}$ & dispersion coefficient of the non-motile bacteria\\
$D_{m}$ & dispersion coefficient of the motile bacteria\\
$\rho$ & fraction of bacteria at the grains\\
$\beta$ & partition coefficient\\
$R$ & retardation factor associated to the convection at the macroscopic scale of the motile bacteria\\
%\bottomrule
    \end{tabular}
\end{center}
\caption{Notation}
    \label{tab:notation}
\end{table}
%%%%%%%%%%%%%%%%%%%%%%%%%%%%%%%%%%%%%%%%%%%%%%%%%%%%%%%%%%%%%%%%

%%%%%%%%%%%%%%%%%%%%%%%%%%%%%%%%%%%%%%%%%%%%%%%%%%%
\section{Theoretical approach}
%%%%%%%%%%%%%%%%%%%%%%%%%%%%%%%%%%%%%%%%%%%%%%%%%%%
We present here the theoretical approach to model the dispersion of
non-motile and motile bacteria. We use the CTRW framework to model the
stochastic motion of bacteria due to pore-scale flow variability and motility, 
based on a spatial Markov model for subsequent particle velocities, 
and a compound Poisson process for motility. This type of approach 
was used to upscale and predict hydrodynamic transport in porous and 
fractured media at the pore and continuum scales~\cite[][]{BS1997, noetinger2016,
  Dentz2018jfm, Hyman2019}. It naturally accounts for the organization
of the flow field along characteristics length scales that are imprinted 
in the host medium. We focus here on the quantification of the streamwise motion
and large scale dispersion of bacteria, which play a key role for the prediction
of the length of bacteria plumes and the distributions of residence times in a
porous medium.

%%%%%%%%%%%%%%%%%%%%%%%%%%%%%%%%%%%%%%%%%%%%5
\subsection{Non-motile bacteria}
\label{sec:nm}

Non-motile bacteria are considered as passive tracer particles that are transported by advection only. Non-motile bacteria move along streamlines of the porescale flow field, and thus explore the porescale velocity spectrum, except for the lowest
velocities close to the grains, due to volume exclusion or molecular
diffusion. Typical trajectories are shown in
Figure~\ref{fig:experiment}. In the following, we model the motion of non-motile bacteria
using a spatial Markov model for particle speeds~\cite[][]{Dentz2016,morales2017,puyguiraud2019upscaling}. 
%%%%%%%%%%%%%%%%%%%%
\subsubsection{Spatial Markov model}
%%%%%%%%%%%%%%%%%%%%
Particle motion is characterized by the spatial persistence of particle velocities over a characteristic length scale, 
which is imprinted in the spatial structure of the porous medium ~\cite[][]{Dentz2016}. This provides a
natural parameterization of bacteria motion in terms of travel
distance. That is, motion is modeled by constant 
space and variable time increments along streamlines. Thus, the
equations of streamwise motion of non-motile bacteria can be written as~\cite[][]{puyguiraud2019upscaling}
\begin{align}
\label{ctrw:nm}
x_{n+1} = x_n + \frac{\Delta s}{\chi}, && t_{n+1} =  t_n + \frac{\Delta s}{v_n},
\end{align}
where $\Delta s$ is the transition length along the tortuous particle
path. The advective tortuosity $\chi$ accounts for streamline
meandering in the pore space between the grains. It quantifies the ratio of the
average streamline length to streamwise distance. Note that this meandering is
different for each streamline and may be correlated to the particle
speed. However, under ergodic flow conditions, the streamline lengths converge
toward the average value, and thus, at scales larger than $\ell_0$ tortuosity
provides a good estimate for the longitudinal displacement. 

The point distribution $p_v(v)$ of particle speeds is given 
in terms of the Eulerian flow speed distribution $p_e(v)$
\begin{align}
\label{pv}
p_v(v) = \frac{v p_e(v)}{v_m}. 
\end{align}
This speed-weighting relation is due to the fact that in this
framework particles make transitions over constant distance, while
the distribution of flow speeds $p_e(v)$ is obtained by measuring 
speeds at constant framerate, this means
isochronically~\cite[][]{Dentz2016, morales2017, puyguiraud2019upscaling}. 
Equations~\eqref{ctrw:nm} constitute a CTRW because bacteria are propagated 
over constant (discrete) distances while time is a continuous variable. In this framework, the position $x(t)$ of a particle at time $t$ is given by 
    $x(t) = x_{n_t}$,
where $n_t = \max(n|t_n \leq t < t_{n+1})$. The displacement moments are defined by $m_i(t) = \langle x(t)^i \rangle$. The displacement variance is given by $\sigma^2(t) = m_2(t) - m_1(t)^2$.

The series $\{v_n\}$ of particle speeds is modeled as a stationary
Markov process whose steady state distribution is given by
Eq.~\eqref{pv}. Specifically, we model $\{v_n\}$ through an Ornstein-Uhlenbeck process
for the unit normal random variable $w_n$ which is obtained from $v_n$
through the transformation~\cite[][]{puyguiraud2019a}
\begin{subequations}
\label{v:process}
\begin{align}
w_n = \Phi^{-1}\left[P_v(v_n)\right], && v_n = P_v^{-1}[\Phi(w_n)],
\end{align}
where $P_v$ is the cumulative speed distribution and $\Phi^{-1}(u)$
the inverse of the cumulative unit Gaussian distribution. The $w_n$
satisfies the Langevin equation
\begin{align}
\label{ou}
w_{n+1} = w_n - \ell_c^{-1} \Delta s w_n + \sqrt{2 \ell_c^{-1} \Delta s} \xi_n,
\end{align}
\end{subequations}
where $\xi_n$ is a unit Gaussian random variable. The length scale $\ell_c$
denotes the characteristic correlation scale of particle speed. It is typically
of the order of the characteristic grain size
$\ell_0$~\cite[][]{puyguiraud2021pore}. However, its exact value
needs to be adjusted from the data for the displacement variance.
The increment $\Delta s$ is chosen such that $\Delta s \ll \ell_c$.
The phase-space particle density $p(x,v,t)$ in this framework is given
by the Boltzmann-type equation~\cite[][]{Comolli2019}
\begin{align}
\label{boltzmann}
\frac{\partial p(x,v,t)}{\partial t} + v \chi^{-1} \frac{\partial
  p(x,v,t)}{\partial x} = - \frac{v}{\Delta s} p(x,v,t) +
\int\limits_0^\infty dv' r(v,\Delta s|v') \frac{v'}{\Delta s}
p(x,v',t),
\end{align}
see also Appendix~\ref{app:ctrwnm}. The initial distribution is given
by $p(x,v, t = 0) = p_0(x,v) = \delta(x)
p_0(v)$, where $p_0(v)$ is the distribution of initial particle
velocities. The propagator, that is, the distribution of particle displacements, is given by
\begin{align}
p(x,t) = \int\limits_0^\infty dv p(x,v,t). 
\end{align}
%
%%%%%%%%%%%%%%%%%%%%%%%%%%%%%
\subsubsection{Asymptotic theory}
%%%%%%%%%%%%%%%%%%%%%%%%%%%%%
The behavior of the upscaled model at travel distances much larger than the correlation length $\ell_c$, can be
obtained by coarse-graining particle motion on a length scale $\ell'_c \geq \ell_c$, such that
\begin{align}
\label{uc:ctrw}
x_{n+1} = x_n + \frac{\ell'_c}{\chi}, && t_{n+1} = t_n + \tau_n,
\end{align}
The transition times $\tau_n = \ell'_c/v_n$ are independent
random variables whose distribution $\psi(t)$ is given in terms of
$p_v(v)$ as
\begin{align}
\label{psi}
\psi(t) = \ell'_c t^{-2} p_v(\ell'_c/t) = \left(\frac{t}{\tau_0}\right)^{-2-\alpha} \frac{\exp(-\tau_0/t)}{\tau_0 \Gamma(\alpha+1)},
\end{align}
where $\tau_0 = \ell'_c/v_0$. $\psi(t)$ is given here by an inverse Gamma distribution because the particle
speed is Gamma-distributed, see Eq.~\eqref{eq:gamma}.  

For the velocity distribution~\eqref{eq:gamma} with $\alpha = 2.25$, the CTRW predicts
asymptotically a Fickian dispersion. That is, for times $t \gg \tau_v$, transport can be
quantified by the advection-dispersion equation~\cite[][]{dentz2003transport}
\begin{align}
\frac{\partial p(x,t)}{\partial t} + u_m \frac{\partial
  p(x,t)}{\partial x} - D_{nm} \frac{\partial^2 p(x,t)}{\partial x^2}
= 0. 
\end{align}
with the average velocity $u_m = v_m
/\chi$ and the dispersion coefficient~\cite[][]{puyguiraud2021pore}
\begin{align}
D_{nm} = \frac{u_m \ell'_c}{2 \chi} \frac{\langle \tau^2 \rangle -\langle \tau \rangle^2}{\langle \tau \rangle^2}. 
\end{align}
The mean and mean squared transition times are defined by 
\begin{align}
\langle \tau^k \rangle = \int\limits_0^\infty dt t^k \psi(t) = \tau_0^k \frac{\Gamma(\alpha + 1 - k)}{\Gamma(\alpha + 1)},
\end{align}
for $k = 1,2$. $\Gamma(\alpha)$ denotes the Gamma function. We find by comparison of the dispersion coefficients from the full spatial Markov model and the CTRW model~\eqref{uc:ctrw} that $\ell_c' \approx 1.57 \ell_c$.

%%%%%%%%%%%%%%%%%%%%%%%%%%%%%%%%%%%%
\subsection{Motile bacteria}
\label{sec:m}
We provide here the theoretical framework to interpret the trajectory data and motion of motile bacteria. 
The motion of motile bacteria is due to advection in the flow field
and their own motility as illustrated in Figure~\ref{fig:experiment}. At zero flow rate, bacteria fluctuate
 in a random walk-like manner characterized by a zero mean displacement with a characteristic
2D projected swimming velocity $v_0 \approx 12 \mu$m/s~\citet{Creppy2019}. At finite flow rate,
bacteria tend to swim along the streamlines, and make excursions
perpendicular to them in order to move toward the solid grains.
Based on the observations of~\citet{Creppy2019} for bacteria motility,
we couple the CTRW model for hydrodynamic transport with a trapping
approach. These authors found that bacteria move towards
the grains at a flow dependent rate $\gamma$ and dwell on the
grain surface for random times $\theta$, which are distributed
according to the trapping time distribution $\psi_f(t)$.  

%%%%%%%%%%%%%%%%%%%%%%%%%%%%%%%%%%%%%%%%
\subsubsection{Spatial Markov model and trapping}

Within the CTRW approach
outlined in the previous section, the trapping of bacteria is represented
by a compound Poisson process for the time $t_n$ of the bacteria after
$n$ CTRW steps. Thus, the equations of motion are given by
\begin{align}
\label{ctrw:m}
x_{n+1} = x_n + \frac{\Delta s}{\chi},
&&
t_{n+1}  = t_n + \frac{\Delta s}{v_n} + \tau(\Delta s/v_n).
\end{align}
for $n > 1$. The initial displacement is $x_0 = 0$ for all bacteria. The initial time is set to $t_0 = 0$. 
The particle speeds $v_n$ evolve according to the process~\eqref{v:process}. The compound trapping time $\tau(r)$ is given by
\begin{align}
\label{cp}
\tau(r) = \sum\limits_{i = 1}^{n_r} \theta_i,
\end{align}
where $\theta_i$ is the trapping time associated to an individual
trapping event, and $n_r$ is the number of trapping events during time $r$. 
 The number of trapping events $n_r$ follows a Poisson process characterized by the rate
$\gamma$, that is, the mean number of trapping events per CTRW step
is $\gamma \Delta s/v_n$. The trapping rate is constant and counts the average
number of trapping events per mobile time. While the trapping properties could
depend for example on the local flow speeds, we use the Poisson process with
constant rate as a robust and simple way of describing the average trapping properties, which
is fully defined by the average number of trapping events per mobile time.
The distribution of compound trapping times
$\tau(r)$, denoted by $\psi_c(t|r)$, can be expressed in Laplace space
by~\cite[][]{Feller:1,Margolin2003}
\begin{align}
\label{psifc}
\psi_c^\ast(\lambda|r) = \exp(-\gamma r [1 - \psi^\ast_f(\lambda)] -\lambda r). 
\end{align}
$\psi_c(t|r)$ denotes the probability that the trapping time is $t$ given that a
trapping event occurred at time $r$. For $n = 1$, we distinguish the proportion $\rho$ of 
bacteria that are initially trapped, and $1 - \rho$ of initially mobile bacteria.
For the trapped bacteria, $x_{1} = 0$ and 
$t_1 = \eta_0$, where the initial trapping time $\eta_0$ 
is distributed according to $\psi_0(t)$. For the mobile bacteria, $x_1$ and
$t_1$ are given by Eq.~\eqref{ctrw:m} for $n = 0$.

We consider here steady state conditions at time $t = 0$. As experimental
trajectories and their starting points are recorded continuously,  it is
reasonable to assume that steady state between mobile and immobile bacteria is
attained. Under steady state conditions, the joint probability of the bacterium
to be trapped and the initial trapping time to be in $[t,t+dt]$ is
\begin{align}
\label{psi0}
\text{P}_0(t) = \int\limits_t^\infty dt' \gamma \exp[-\gamma (t' - t)] \psi_f(t')
\end{align}
see Appendix~\ref{app:psi0}. The trapping times are assumed to be exponentially
distributed, that is,
\begin{align}
\label{psif}
\psi_f(t) = \exp(-t/\tau_c)/\tau_c
\end{align}
with $\tau_c$ the characteristic trapping time. This means, we use
 Poissonian statistics to account for the effective retention
of motile bacteria in the vicinity of grain surfaces. This picture is
classically based on the idea that
the run to tumble process promoting surface detachment is itself a
memory-less Poisson process \cite[][]{berg2018random}. However,  there
has been recent evidence that the  run-time distribution for
bacterial motion in a free fluid is a long-tail non-Poissonian process
\cite[][]{figueroa20203d},  which is also at the origin of a long-tailed
distributions of bacteria sojourn times at flat surfaces
\cite[][]{Junot2022}. For porous media, there are currently no direct measurements that
offer a quantitative microscopic description of the complex exchange
processes taking place between the surface regions and the flowing
regions. Thus, we adopt Poissonian statistics characterized by the
mean retention time $\tau_c$ as a model of minimal assumptions. We hope that
our conceptual approach, which provides a model of the emerging
transport process, will motivate more detailed experimental
investigations on this central question.
% This is a reasonable assumption
% in the absence of experimental evidence on the form of $\psi_f(t)$. It is
% supported by classical models for the distribution
% of run times between tumbling events~\cite[][]{berg2018random}.
% The exponential assumption implies that the mass
% transfer process follows linear first-order kinetics characterized by the trapping rate
% $\gamma$ and the release rate $1/\tau_c$.
Using Eq.~\eqref{psif} in
Eq.~\eqref{psi0}, we obtain
\begin{align}
\text{P}_0(t) = \frac{\beta}{1 + \beta} \frac{\exp(-t/\tau_c)}{\tau_c},
\end{align}
where we define the partition coefficient $\beta = \gamma \tau_c$. Thus, the
fraction of trapped bacteria is $\rho = \beta/(1 + \beta)$, and the initial
trapping time distribution is $\psi_0(t) = \psi_f(t)$. Thus, the steady state
partitioning of bacteria is directly related to their motility through the
trapping rate $\gamma$ and mean dwelling time $\tau_c$ on the grain surface. 

Note that this picture does not account for the tortuous particle path
on the grain surfaces, which is represented as a localization event
at fixed positions. Grain-scale bacteria motility could eventually be modeled by an additional process. However, here we focus on large scale
bacteria dispersion and only account for tortuosity due to the flow
path geometry. As above the bacteria position $x(t)$ at time $t$ is given by $x(t) = x_{n_t}$. The expressions for the displacement moments and variance are analogous. 

The density $p_s(x,v,t)$ of mobile bacteria in the stream is quantified by the
non-local Boltzmann equation
\begin{align}
\frac{\partial p_s(x,v,t)}{\partial t} &+ \frac{\partial}{\partial t}
\int\limits_0^t dt' \gamma \phi(t - t') p_s(x,t') +
\frac{v}{\chi} \frac{\partial
  p_s(x,v,t)}{\partial x} =
\nonumber\\
& \rho \delta(x) p_0(v) \psi_f(t) -\frac{v}{\Delta s} p_s(x,v,t) +
\int\limits_0^\infty dv' r(v|v') \frac{v'}{\Delta s} p_s(x,v',t'), 
\label{eq:boltzmann}
\end{align}
see Appendix~\ref{app:ctrwm}. We defined
\begin{align}
\phi(t) = \int\limits_t^\infty dt' \psi_f(t'), 
\end{align}
the probability that the trapping time is larger than $t$. Equation~\eqref{eq:boltzmann} reads as follows. The
evolution of the particle density in the stream is given by
(second term on the left side) particle exchange between the stream and
grain surface, (third term on the left) advection by the local velocity,
(first term on the right side) release of bacteria that were initially
on the grains, (second and third terms on the right) velocity transitions along the trajectory. 

The total bacteria density is given by
\begin{align}
\label{p}
p(x,v,t) = p_s(x,v,t) + p_g(x,v,t).
\end{align}
The density $p_g(x,v,t)$ of bacteria on the grains is given by 
\begin{align}
\label{pgps}
p_g(x,v,t) = \int\limits_0^t dt' \phi(t - t') \gamma p_s(x,v,t') +
\delta(x) \rho \phi(t) p_0(v)
\end{align}
This first term on the right side reads as follows. The density of bacteria on the grains
is given by the probability per time $\gamma p_s(x,t')$ that bacteria
are trapped at time $t'$ times the probability $\phi(t - t')$ that
the  trapping time is longer than $t - t'$. The second term denotes
the bacteria that are initially trapped and whose trapping time is
larger than $t$. The speed $v$ associated
with a bacterium on the grain should be understood as the bacteria
speed before the trapping events.

%%%%%%%%%%%%%%%%%%%%%%%%%%%%%%%%%%%%%%%%%%%%%%
\subsubsection{Asymptotic theory\label{sec:asymptoticm}}
%%%%%%%%%%%%%%%%%%%%%%%%%%%%%%%%%%%%%%%%%%%%%%

Similar to the discussion in the previous section for the non-motile
bacteria, for distances much larger than $\ell_c$, particle motion can
be coarse-grained such that
\begin{align}
x_{n+1} = x_n + \ell'_c, && t_{n+1}  = t_n + \tau_n+ \tau(\tau_n),
\end{align}
where the advective transition times $\tau_n = \ell'_c/v_n$ are distributed according to
Eq.~\eqref{psi}. $\tau(r)$ describes the compound Poisson process
defined above. The propagator $p_s(x,t)$ of bacteria in the stream
for this equation of motion is quantified by the non-local
advection-dispersion equation
\begin{align}
\frac{\partial p_s(x,t)}{\partial t} &+ \frac{\partial}{\partial t}
\int\limits_0^t dt' \gamma \phi(t - t') p_s(x,t') \nonumber\\
& + u_m
\frac{\partial p_s(x,t)}{\partial x} - D_{nm} \frac{\partial^2
  p_s(x,t)}{\partial x^2} = \rho \delta(x) \psi_f(t),
\label{eq:mrmt}
\end{align}
while the distribution $p_g(x,t)$ of bacteria at the grains is given
by
\begin{align}
\label{pgmrmt}
p_g(x,t) = \int\limits_0^t dt' \phi(t - t') \gamma p_s(x,t') +
\delta(x) \rho \phi(t).
\end{align}

Asymptotically this means for times $t \gg \tau_c$, the transport of
the bacteria concentration $p(x,t)$ can be described by the
advection-dispersion equation
\begin{align}
\label{ADE:pm}
\frac{\partial p(x,t)}{\partial t} + \frac{u_m}{R}
\frac{\partial p_s(x,t)}{\partial x} - D_{m} \frac{\partial^2
  p_s(x,t)}{\partial x^2} = 0,
\end{align}
see Appendix~\ref{app:asymptotic}. The retardation coefficient $R$ and
the asymptotic dispersion coefficient $D_{m}$ are given by the
explicit expressions
\begin{align}
\label{eq:R}
R &= 1 + \gamma \tau_c =\frac{1}{1-\rho},
\\
D_m &= D_{nm} (1 - \rho) + u_m^2 \tau_c \rho (1-\rho)^2.
%D^\ast = \frac{D}{R} + \frac{\overline u^2 \tau_c (R-1)}{R^2}. 
\label{eq:D}
\end{align}
By definition, $R$ compares the average velocity of motile bacteria
with the average flow velocity. In absence of trapping, $\rho=0$ and
$R=1$, the bacteria are transported in the porous media with an
average velocity equal to the average fluid velocity. If trapping is
present, retardation increases, indicating a decrease of the
average bacteria velocity compared to the fluid velocity. The retardation
coefficient is directly related to bacterial motility, which in our modeling
framework is expressed by the trapping rate $\gamma$ and the mean retention time
$\tau_c$ for which a bacterium dwells at the grain surface.

%The dispersion coefficient, shows a non-monotonic behavior as a
%function of $\rho = \gamma \tau_c$ due to the second term on the right side, which quantifies the interaction between bacteria motility and hydrodynamic
%transport. Dispersion of bacteria is directly related to their motility through the trapping rate $\gamma$ and trapping time $\tau_c$. 
The asymptotic dispersion coefficient in Eq.~\eqref{eq:D} contains
two terms. The first term $D_{nm} (1-\rho)$ corresponds to the
so-called dispersion coefficient at steady state
\citep{Yates1988,Tufenkji2007}. It predicts a reduction of the
dispersion coefficient of the motile bacteria compared to the
non-motile concomitant with the reduction of the average velocity of
the bacteria population. It accounts for the dispersion of the motile
proportion $1-\rho$ only. The second term quantifies a mechanism similar to the
Taylor dispersion. It originates from the spread of the bacteria plume due to
fast transport in the pores and localization at the grains. The resulting
dispersion effect can be rationalized as follows. The typical separation
distance between localized and mobile bacteria, that is, the dispersion length
is $u_m \tau_c$, while the dispersion time is $\tau_c$. The corresponding
dispersion coefficient is dispersion length squared divided by
dispersion time, which gives exactly the scaling $u_m^2 \tau_c$ of
Eq.~\eqref{eq:D}. As we will see in the next section, this interaction can
lead to a significant increase of bacteria dispersion compared to
non-motile bacteria.

Asymptotic bacteria transport is predicted to obey the advection-dispersion
equation with constant parameters for two reasons. First, the distribution of
particle velocities does not tail towards low values, that is, mean and mean
squared transition times are finite. Second, the distribution of retention times
is exponential. Thus, for times large compared to the characteristic mass
transfer times, the support scale can be considered as well-mixed, and, similar
to Taylor dispersion~\citep{Taylor:1953}, and generalized Taylor dispersion
\citep{brenner}, transport can be described by an advection-dispersion equation.   

%%%%%%%%%%%%%%%%%%%%%%%%%%%%%%%%%%%%%%%%%
\section{Results}
We discuss the experimental results for the displacement means and
variances, as well as the displacement distributions, in the light of the theory
presented in the previous section. As discussed in
Appendix~\ref{app:tracks}, the number of experimentally observed
tracks decreases with the travel time, which introduces a bias toward
slower bacteria. Thus, in the following, we consider travel times
shorter than $5 \tau_v$ in order to avoid a too strong bias toward
slow bacteria. Even so, as we will see below, there is a slowing down
of the mean displacement with increasing travel time, specifically for the
motile bacteria. 

The proposed theoretical approach for the non-motile bacteria has one parameter
that needs to be adjusted, the correlation scale $\ell_c$, which typically is of
the order of the grain size. It is adjusted here from the data for the
displacement variance for the non-motile bacteria. The approach for the motile bacteria has two additional parameters, the
trapping rate $\gamma$ and the mean trapping time $\tau_c$.
The partition coefficient $\beta = \gamma \tau_c$ is adjusted from the mean
displacement data for the motile bacteria, while the trapping time $\tau_c$ is
adjusted from the data for the displacement variance of the motile bacteria.
Thus, the non-motile CTRW model needs to adjust one parameter, which is of the order of the grain size.
The motile CTRW model needs to adjust two parameters, which are
related to the partitioning of bacteria between flowing and stagnant regions
close to the grains. 

%%%%%%%%%%%%%%%%%%%%%%%%%%%%%%%%%%%%%%%%%
\subsection{Dispersion of non-motile bacteria}

%%%%%%%%%%%%%%%%%%%%%%%%%%%%%%%%%%%%%%%%%%%%%%%
\begin{figure}
  \centering
\includegraphics[width=0.45\textwidth]{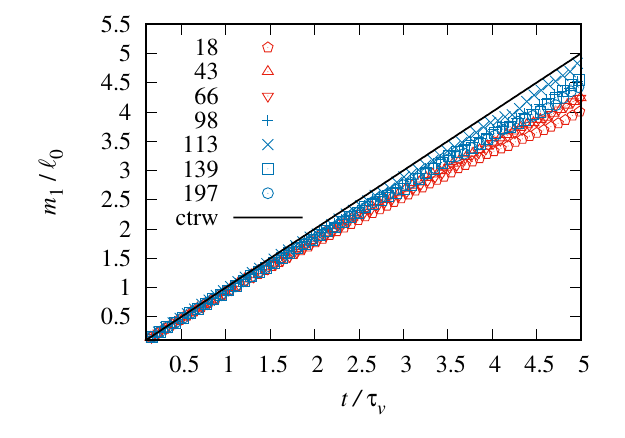}
\includegraphics[width=0.45\textwidth]{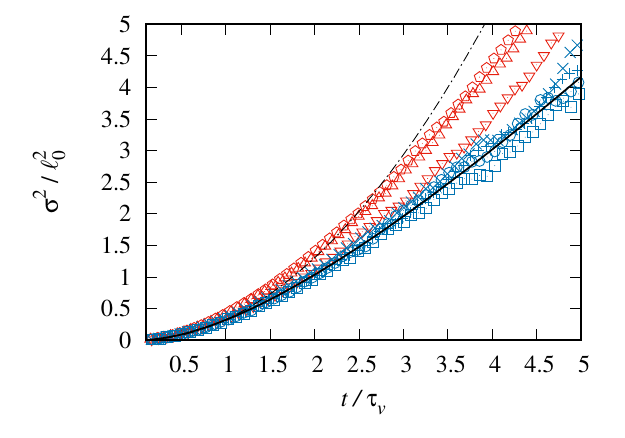}

  \caption{(Left panel) Normalized mean displacements and (right panel)
    normalized displacement variances for non-motile bacteria as a function of normalized time. The solid lines
    denote the estimate from the CTRW model. The dash-dotted line in
    the right panel indicates the initial ballistic growth. }
  \label{fig:moments:nm}
\end{figure}
%%%%%%%%%%%%%%%%%%%%%%%%%%%%%%%%%%%%%%%%%%%%%%% 

%%%%%%%%%%%%%%%%%%%%%%%%%%%%%%%%%%%%%%%%%%%%%%%
\begin{figure}
  \centering
\includegraphics[width=0.45\textwidth]{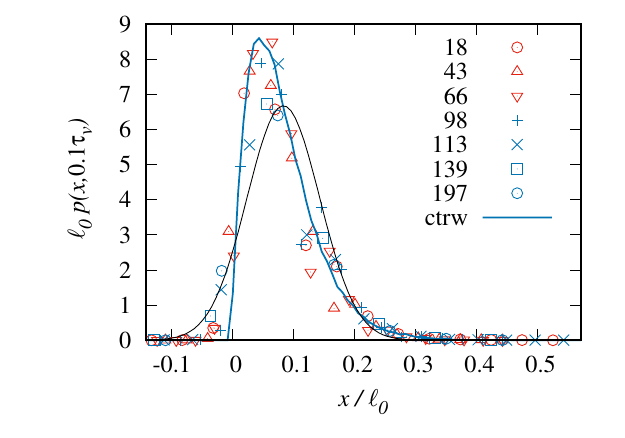}
\includegraphics[width=0.45\textwidth]{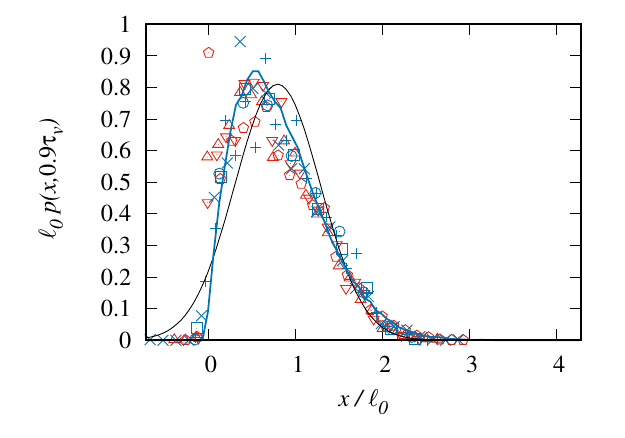}
\includegraphics[width=0.45\textwidth]{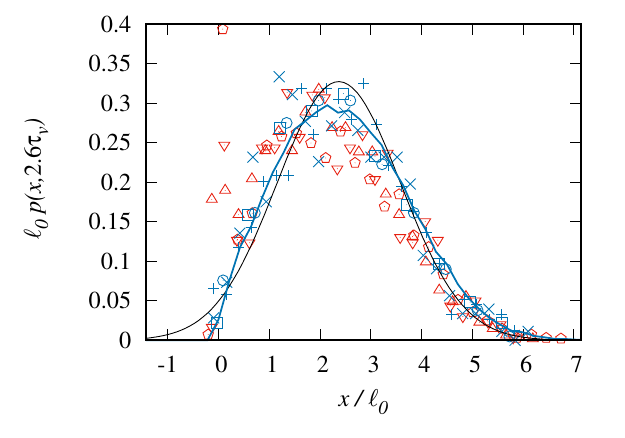}
\includegraphics[width=0.45\textwidth]{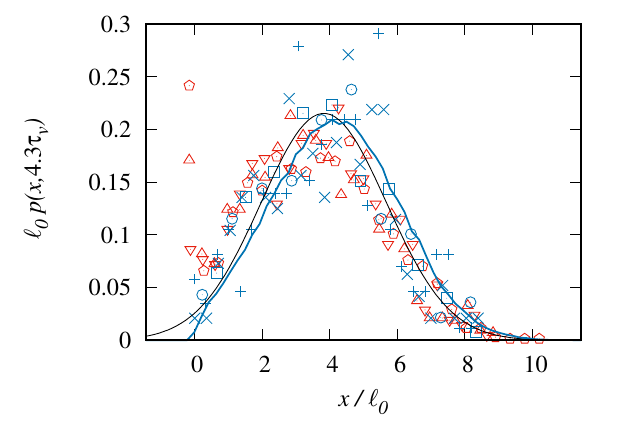}
\caption{Propagators of non-motile bacteria at (left to
  right) $t = 0.1,0.9,2.6,4.3 \tau_v$. The blue solid lines denote the
  prediction of the CTRW model, the black lines are the fits from the
  Gaussian transport model that is characterized by the corresponding
  measured displacement mean and variance shown in
  Figure~\ref{fig:moments:nm}.}
  \label{fig:data:nm}
\end{figure}
%%%%%%%%%%%%%%%%%%%%%%%%%%%%%%%%%%%%%%%%%%%%%%% 

Figures~\ref{fig:moments:nm} and~\ref{fig:data:nm} show
 displacement means and variances and the propagators for non-motile
bacteria at different flow rates and for the same dimensionless
times. Time is non-dimensionalized by the mean advection time over
the size of a grain, which implies that the propagators are reported for
the same mean travel distances. The CTRW model uses the velocity
distribution~\eqref{eq:gamma} with $\alpha = 2.25$, the correlation
length $\ell_c \approx 2 \ell_0$ and the advective tortuosity $\chi =
1.2$.

The mean displacement is linear with a slightly higher slope at short
than at large times. It starts deviating from the expected behavior $m_1(t) =
u_m t$ at around $t = 2 \tau_v$. We relate this behavior to a bias due to the decrease in
the number of tracks as discussed in Appendix~\ref{app:tracks}. The
displacement variance shows a ballistic behavior at $t < \tau_v$, this
means it increases as $t^2$. Then for $t > \tau_v$ it increases
superlinearly, which can be seen as a long cross-over to normal behavior. 
These behavior are accounted for by the CTRW model. For flow
velocities $u_m \leq 66\mu$m/s, we observe a larger variance than
for the higher flow rates. This, and the slightly smaller mean
displacements compared to higher flow rates, can be attributed to the
localization of some bacteria at the origin (see Figure~\ref{fig:data:nm}),
which causes a chromatographic dispersion effect, which is discussed in more
detail for the motile bacteria.  

Figures~\ref{fig:data:nm} compares the experimental data for the propagators
with the results of the CTRW model. The propagators are asymmetric but compact,
meaning that there is no significant forward or backward tails in the
distribution. For comparison, we plot a
Gaussian shaped propagator characterized by the mean displacement and
displacement variance shown in Figure~\ref{fig:moments:nm}.
The asymmetry decreases with increasing travel time and the
propagators become closer to the corresponding Gaussian.
The CTRW model captures the initial asymmetry and
the transition to symmetric Gaussian behavior for all
flow rates. 

%%%%%%%%%%%%%%%%%%%%%%%%%%%%%%%%%%%%%%%%%
\subsection{Dispersion of motile bacteria}
%%%%%%%%%%%%%%%%%%%%%%%%%%%%%%%%%%%%%%%%%

% We
% investigate the interaction of hydrodynamic transport along
% streamlines and motility on the dispersion of bacteria, by comparing
% the transport behavior of motile and non-motile bacteria at different
% flow rates, and derive an upscaled stochastic transport model based on the CTRW
% for non-motile bacteria presented in the previous section.

%%%%%%%%%%%%%%%%%%%%%%%%%%%%%%%%%%%%%%%%%%%%%%%
\begin{figure}
  \centering
\includegraphics[width=0.46\textwidth]{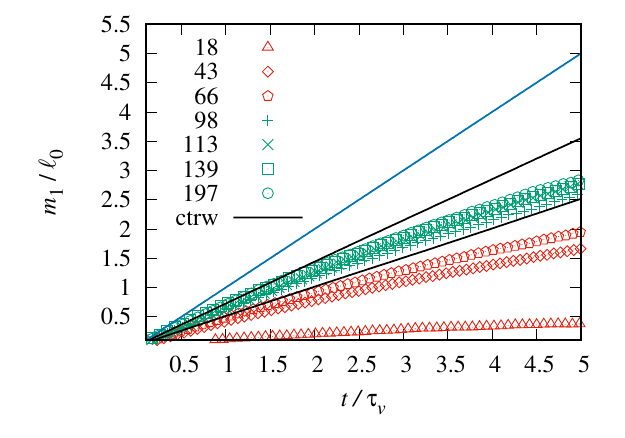}
\includegraphics[width=0.46\textwidth]{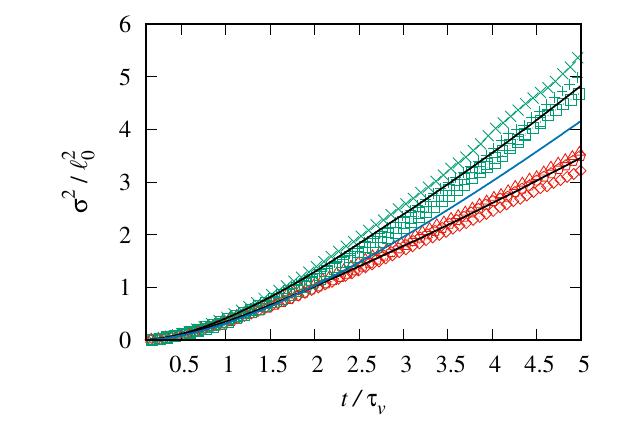}

  \caption{(Left panel) Normalized mean displacements and (right panel)
    normalized displacement variances for motile bacteria, as a function of normalized time. The experimental data
    are denoted by the symbols, the corresponding CTRW model results
    by the thick solid lines. The CTRW model uses $\tau_c = 2.5
    \tau_v$ and $\beta = \gamma \tau_c = 0.4$ for $u_m \geq 98\mu$m/s and
    $\tau_c = 2 \tau_v$ and $\beta = 1$ for $u_m = 66
    \mu$m/s. The solid blue lines denote the model outcomes for the
    non-motile bacteria.}
  \label{fig:moments}
\end{figure}
%%%%%%%%%%%%%%%%%%%%%%%%%%%%%%%%%%%%%%%%%%%%%%% 

\begin{figure}
  \centering
\includegraphics[width=0.46\textwidth]{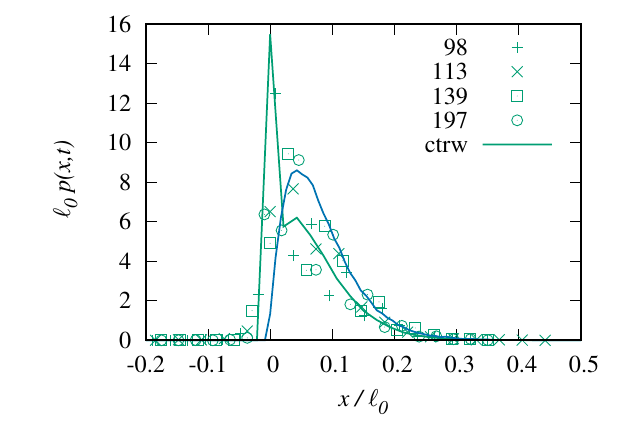}
\includegraphics[width=0.46\textwidth]{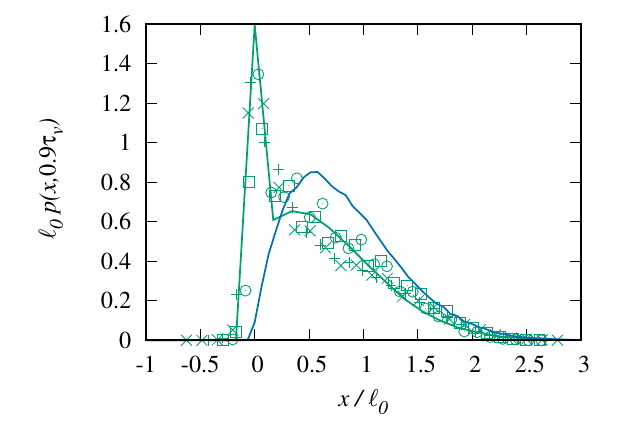}

\includegraphics[width=0.46\textwidth]{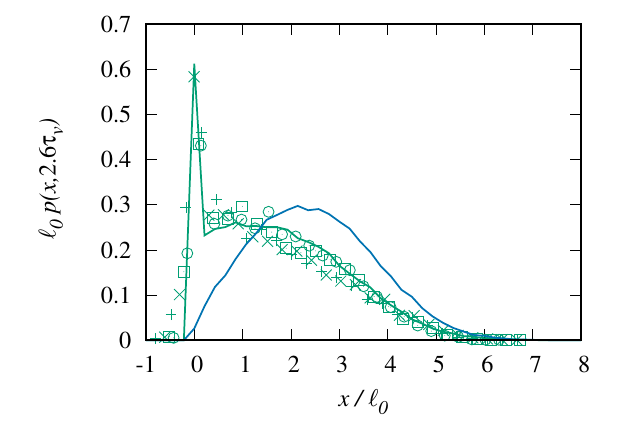}
\includegraphics[width=0.46\textwidth]{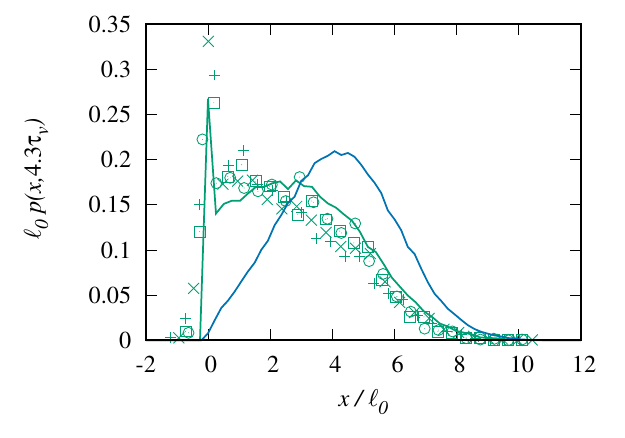}
\caption{Distributions of motile bacteria for high flow
    rates at (left to right)
$t = 0.1,0.9,2.6,4.3 \tau_v$. The blue solid lines denote the
    corresponding predictions of the CTRW model for the non-motile bacteria.\label{fig:data}}
\end{figure}

\begin{figure}
\includegraphics[width=0.46\textwidth]{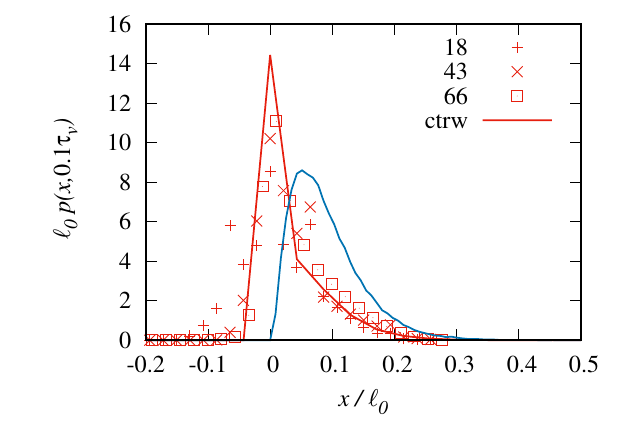}
\includegraphics[width=0.46\textwidth]{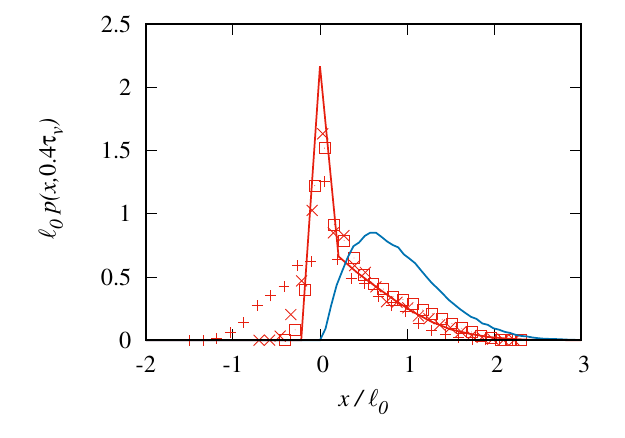}

\includegraphics[width=0.46\textwidth]{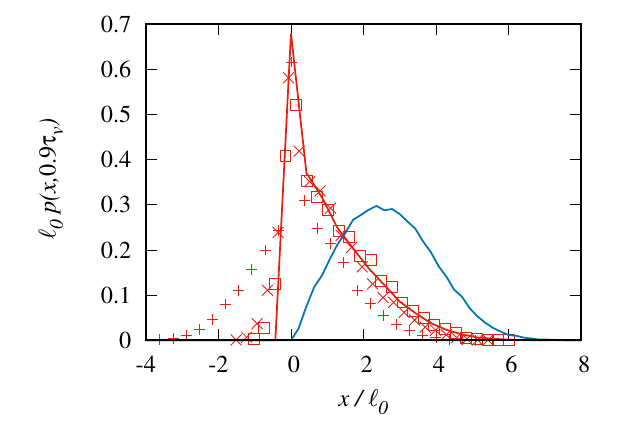}
\includegraphics[width=0.46\textwidth]{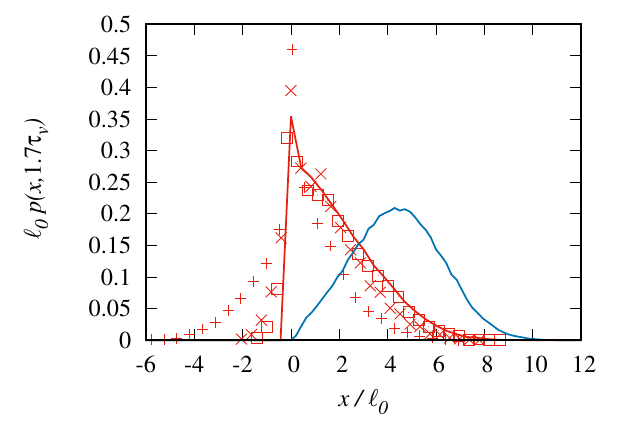}
  \caption{Propagators of motile bacteria for low flow rates at (left to right)
$t = 0.1,0.9,2.6,4.3 \tau_v$. The blue solid lines denote the
    prediction of the CTRW model for the non-motile bacteria.}
  \label{fig:datalflowrate}
\end{figure}

Figures~\ref{fig:moments},~\ref{fig:data} and~\ref{fig:datalflowrate} show the displacement
mean and variance, and the propagators
for the motile bacteria at different
flow rates. As in the previous section, time is measured in units of
$\tau_v$, that is, it measures the mean number of grains the bacteria have
passed. The propagators are measured at the same non-dimensional
times, that is, at the same mean distance. The motile CTRW model is
parameterized by the same correlation length and tortuosity as the
non-motile model. The partition coefficient $\beta = \gamma \tau_c$ is adjusted from the
early time behavior of the mean displacement, which is predicted to behave as
\begin{align}
\label{m1R}
m_1(t) = \frac{u_m t}{R} = \frac{u_m t}{1 + \beta},
\end{align}
because we consider the system to be initially in a steady state. The
characteristic trapping time is adjusted from the displacement
variance by keeping $\beta$ fixed.  We adjust $\tau_c = 2.5 \tau_v$ and $\beta =
\gamma \tau_c = 0.4$ for $u_m \geq 98 \mu$m/s, and $\tau_c = 2 \tau_v$ and
$\beta = 1$ for $u_m = 66\mu$m/s. 

As shown in Figure~\ref{fig:moments}, the mean displacement
is consistently lower for the motile than for the non-motile
bacteria, which is due to migration toward the grain surfaces and
localization at the grains. The means displacement initially evolves
linearly until a time of about $2 \tau_v$ and from there, the evolution
slows down. We relate this to the decrease of the number
of experimentally observed  tracks, which induces a bias toward slow tracks as
discussed in Appendix~\ref{app:tracks}. In contrast to the mean
displacement, the displacement variance can be larger than its
non-motile counterpart for $u_m \geq 98 \mu$m/s and lower for $u_m
\leq 66 \mu$m/s. The data seem to fall into two groups for high and
low flow rates, except for $u_m = 18\mu$m/s. In this case, the flow
velocity is of the order of the swimming velocity $v_0 \approx
12\mu$m/s. The data indicates that that the density of trapped particles is
higher at high than at low flow rates. The possible mechanisms for these
behaviors are discussed in Section~\ref{sec:discussion}.  

These behaviors are also reflected in the propagators shown in
Figure~\ref{fig:data} for high flow rates with $u_m \geq 98 \mu$m/s and in
Figure~\ref{fig:datalflowrate} for $u_m \leq 66\mu$m/s. 
The green symbols in Figure~\ref{fig:data} denote the experimental data rescaled by the
mean grain size $\ell_0$, the solid green lines, the corresponding solution from
the CTRW model for the parameters $\tau_c = 2.5 \tau_v$ and $\beta = \gamma
\tau_c = 0.4$. Analogously, the red symbols in Figure~\ref{fig:datalflowrate}
denote the experimental data rescaled by the
mean grain size $\ell_0$. The solid red lines show the corresponding solution from
the CTRW model for the parameters $\tau_c = 2 \tau_v$ and $\beta = 1$.
For comparison, we also plot the corresponding CTRW solution for
the non-motile bacteria, marked by the blue solid lines.
The motile propagators are delayed compared to the
non-motile bacteria. They are characterized by
a localized peak around zero and a pronounced forward tail, 
which can be attributed (i) to slow motion towards and around grains and (ii) to
fast motion in the main pore channels. Figure~\ref{fig:data} shows that the
propagators at high flow rates ($u_m \geq 98\mu$m/s) overlap, which indicates
that bacteria motion scales with the mean flow. Similarly, for the low flow
rates ($u_m \leq 66\mu$m/s) shown in Figure~\ref{fig:datalflowrate}, we observe
overlap in the forward tails, which are advection-dominated due to transport in the
pore-channels. However, the upstream tails that develop starting from the localized
peak do not group together. They can be attributed to bacteria
motility, which is independent of the flow rate. This is most
pronounced for $u_m = 18\mu$m/s, which is characterized by strong
localization and an almost symmetric propagator. 
The features of peak localization and forward tailing show that steady state in
the macroscopic transport behavior has not been attained at the largest
observation time. At asymptotic times, that is, for $t \gg \tau_c$, the theoretical
model given by Eq.~\eqref{ADE:pm} predicts Fickian transport characterized by
symmetric propagators.

The data for the displacement moments and propagators seems to indicate that the
data are grouped in two families, which we have highlighted by using two different
colors. These observations are in agreement with the behaviors of the
speed PDFs shown in Figure~\ref{fig:vpdf}. We therefore fit each family
separately. From the early time evolution of the mean displacements, we
adjust the partition coefficient $\beta = 0.4$ for $u_m \geq
98\mu$m/s, $\beta = 1$ for $u_m \leq 66\mu$m/s.  
% $\beta = 1.25$ for $u_m =
% 43\mu$m/s and $\beta = 6.25$ for $u_m = 18 \mu$m/s.
% In the following, we focus on the data for $u_m \geq 98\mu$m/s and $u_m = 66\mu$m/s to
% highlight the differences between high an low flow rates.
For $u_m \geq 98\mu$m/s, we adjust from the displacement data $\tau_c = 2.9
\tau_v$ and for $u_m = 66\mu$m/s, we adjust $\tau_c = 2.3 \tau_v$.  
 
%%%%%%%%%%%%%%%%%%%%%%%%%%%%%%%%%%%%%%%%%%%%%%%
%%%%%%%%%%%%%%%%%%%%%%%%%%%%%%%%%%%%%%%%%%%%%%%
 
With these parameter sets, the CTRW model is able to
describe the propagators and displacement moments as shown in
Figures~\ref{fig:moments} and~\ref{fig:data}. For the lowest flow velocity, bacteria are able to
swim upstream over relatively long distances. The subsequent backward
tail that develops because of the upstream motion is clearly visible
in Figure \ref{fig:data} (bottom row), and also, to a smaller extent
at the higher flow rates (top row). This effect is not accounted for
in the model that assumes that the trapping is localized and that
trapped bacteria do not move once trapped. 

Since $\tau_v \propto 1/u_m$, our results indicate that the trapping
rate increases linearly with the average flow velocity $u_m$ while the
characteristic trapping time decreases linearly with
$u_m$. We used different values for $\beta = \gamma \tau_v$ and $\tau_c$ to adjust
the two sets. Recall that the fraction of trapped bacteria
$\rho$ is $\beta/(1+\beta)$. Each set thus corresponds
to a different value of the fraction of trapped bacteria. The fraction of
trapped bacteria is high at low velocities ($\rho \geq 0.5$) and
decreases towards an asymptotic value of about $\rho = 0.3$ as the
flow velocity is increased. The fraction of trapped bacteria is also related to the
retardation coefficient $R$ through Eq.~\eqref{eq:R}, which is estimated
from the experimental data for the mean bacteria displacement according to
relation~\eqref{m1R}. The dependence of $R$ and thus $\rho$ on the flow rate is
further discussed in the next section.   

%%%%%%%%%%%%%%%%%%%%%%%%%%%%%%%%%%%%%%%%%%%%%%%
%%%%%%%%%%%%%%%%%%%%%%%%%%%%%%%%%%%%%%%%%%%%%%% 

%%%%%%%%%%%%%%%%%%%%%%%%%%%%%%
\subsection{Asymptotic dispersion and retardation}
\label{sec:macro-m}
\begin{figure}
  \centering
\includegraphics[width=0.46\textwidth]{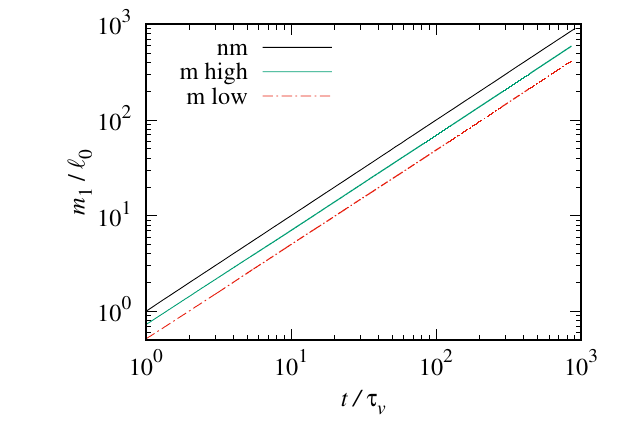}
\includegraphics[width=0.46\textwidth]{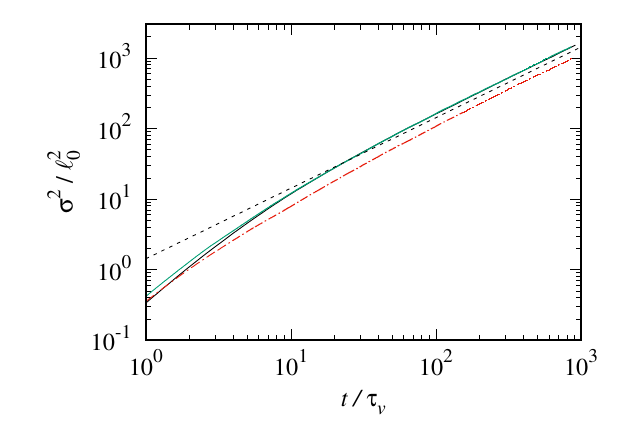}

\includegraphics[width=0.46\textwidth]{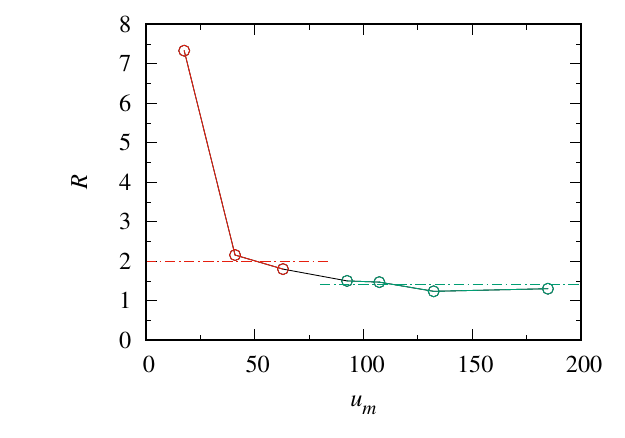}
\includegraphics[width=0.46\textwidth]{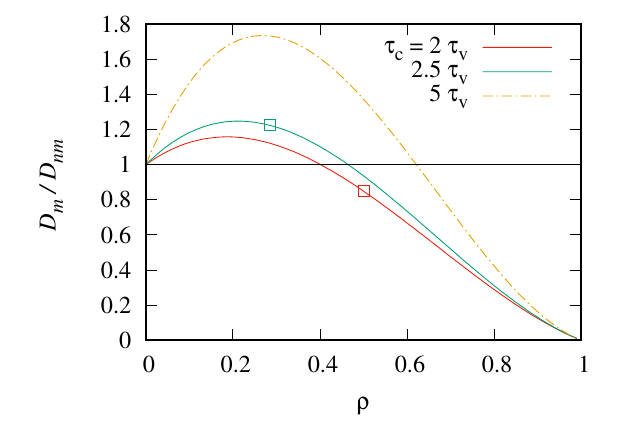}

\caption{Top panel: Model predictions for the displacement (left) means and
 (right) variances of motile and non-motile bacteria. Bottom panel
 left: retardation coefficient from experimental data. The dash-dotted
 line indicates the values used in the CTRW model at (green) high and
 (red) low flow rates. Bottom panel right: Dispersion coefficient for
 the motile bacteria as a function of the fraction $\rho$ of trapped bacteria
    for $\tau_c = 2 \tau_v, 2.5 \tau_v, 5 \tau_v$. The squares denote the
   dispersion coefficient at the $\rho$-values for (green square) $u_m \geq
   98\mu$m/s and (red square) $u_m = 66\mu$m/s.} 
  \label{fig:moments:a}
\end{figure}

The CTRW model allows to extrapolate the transport behaviors to times
that cannot be reached in the experiment. The top panels of
Figure~\ref{fig:moments:a} show the displacement mean and variance up
to times of $1000 \tau_v$. We see that both observables evolve
linearly at asymptotic times. The mean displacement indicates a lower
average velocity for the motile than for the non-motile bacteria, which is due
to trapping. The displacement variance on the other hand is larger for
the motile than for the non-motile at high flow rates, which indicates
stronger motile dispersion. This effect can be quantified by Eqs.~\eqref{eq:R}
and~\eqref{eq:D} for the retardation coefficient and asymptotic dispersion
coefficient.

The retardation coefficient $R = 1/(1-\rho) = 1+ \beta$ can be estimated directly from the
experimental data for the mean displacement according to Eq.~\eqref{m1R}. The
left panel at the bottom of Figure \ref{fig:moments:a} shows that the
retardation coefficient decreases with increasing flow rate, which is consistent
with the values adjusted for $\beta$ in the previous section. Thus the data
shows also that the fraction $\rho$ of trapped particles decreases with
increasing flow rate. 

The behavior of $D_m$ as a function of the proportion $\rho$ of trapped
bacteria is shown in the bottom panel of
Figure~\ref{fig:moments:a}. The solid line shows the theoretical
behavior of $D_{nm}$ for $\tau_c = 2.5\tau_v$ and $\tau_c = 2 \tau_v$,
which corresponds to the value used in the CTRW model. The green and red symbols denote the
values obtained from the CTRW models at high and low flow rates. We
see that at low fractions of immobile bacteria, the Taylor term in
Eq.~\eqref{eq:D} dominates and motile bacteria disperse more than
non-motile. At high proportions of trapped bacteria, localization
dominates over the Taylor mechanism, and motile dispersion is lower
than non-motile. Figure~\eqref{fig:moments:a} illustrates the
competition between the trapping time $\tau_c$ and the proportion
$\rho$ of trapped bacteria. For increasing $\tau_c$, motile dispersion can be
significantly larger than non-motile dispersion. 

%%%%%%%%%%%%%%%%%%%%%%%%%%%%%%%%%%%%%%%%%%%%%%%
%%%%%%%%%%%%%%%%%%%%%%%%%%%%%%%%%%%%%%%%%%%%%%% 

%%%%%%%%%%%%%%%%%%%%%%%%%%%%%%%%%%%%%%%%%%%%
\section{Discussion\label{sec:discussion}}
%%%%%%%%%%%%%%%%%%%%%%%%%%%%%%%%%%%%%%%%%%%%
% We consider scenarios,  
% biological processes (predation, death, growth) and
% chemical absorption/desorption on the grains are not considered.
We study the interaction between bacteria motility and flow variability, and its impact on
the dispersion of bacteria. % We disregard biological processes (predation, death, growth) and
% chemical absorption/desorption on the grains.
To do so we use data obtained in a microfluidic chip containing randomly
placed obstacles, in which thousands of non-motile and motile bacteria were tracked at different flow rates.
This geometry reproduces the structure of a
porous medium on the scale of a few pores, and is thus ideal to study transport phenomena at the
pore scale. Because bacteria do not adhere to the surface of
the flow cell, this setup allows to study the first step of
filtration which consists of the transport of bacteria from the
flowing fluid to regions of low flow in the vicinity of solid grains.

Bacteria motion is quantified by a CTRW approach that is based on a Markov model for
equidistant particle speeds. The experimental data for the displacement of non-motile bacteria is used to
constrain the velocity correlation length, which is of the order of the grain size.
Bacteria motility is modeled in this framework by a trapping process, which accounts for the rheotactic motion
toward and along the grain surfaces by a trapping rate $\gamma$ and characteristic dwelling time $\tau_c$.  
The ratio between trapped and mobile bacteria at steady state is measured by the partition coefficient
$\beta = \gamma \tau_c$.

Adjustment of the model to the experimental data reveals two main features. Firstly, we observe
that $\gamma \propto u_m$ and $\tau_c \propto 1/u_m$.
The increase of the trapping rate with the flow rate can be explained by the constant reorientation of the
bacteria by the flow. The frequency by which bacteria point toward the grains increases with the flow rate,
which may explain the increase of the trapping rate. Similarly, for increasing flow rate, shear increases on the grains and thus the
area for motion around the grains decreases and the bacteria are more
easily blown off by the flow. This can explain why the residence time
decreases with flow rate. A model that supports this idea is proposed
in Appendix \ref{app:arrachement}.

Secondly, we observe that the ratio $\beta$ between trapped and mobile bacteria is different at high and low
flow rates. This observation indicates a transition between a regime at low flow rates,
where motility favors trapping with a high density of trapped
bacteria (about $50\%$ of trapped bacteria), to a regime at high flow rates, where the flow hinders trapping
(about $30\%$ of trapped bacteria). Two phenomena may
contribute to this change. The first comes from the volume of fluid in
which the bacteria can be considered as trapped. This fraction can be
separated in two: a part where the velocity is very small  (this part
corresponds to the dark blue regions that can be seen in
Figure \ref{fig:experiment} and is always present for all the flow rates
used) and a second contribution which comes from the regions of flowing
fluid where the average flow velocity is less than the swimming
velocity. In those volumes, which are located close to the grain
surface the bacteria trajectories are little influenced by the flow and they swim much like in a quiescent fluid.
Bacteria can be considered trapped when they swim along the grain surface. This contribution however decreases with the flow rate
reducing in turn the density of trapped bacteria as observed. The
second contribution comes from the diffusion due to the constant
reorientation of the bacteria. In a fluid at rest, the trajectories of
the bacteria can be decomposed as a succession of runs followed by
tumbles that reorient the bacteria. At large scale, the
reorientation is diffusive and can be characterized by the translational 
diffusion coefficient $D_b$. For E.coli we have here $D_b \approx 243\mu$m$^2$/s~\citep{Creppy2019}. 
In a shear flow, bacteria constantly tumble and
are reoriented at a frequency set be the shear rate $\dot{\gamma}$
\citep{Jeffery1922}. When the P\'eclet number defined as
$Pe = u_m \ell_0/{2 D_b}$ is of the order of $1$. For a grain size of $\ell_0 = 30\mu$m, 
we have  $Pe \simeq u_m/(16 \mu$m/s). Random
orientation will thus dominate shear alignment
for the lowest flow rate with little or no influence at high flow velocity.

\section{Conclusions}

In conclusion, to understand the dispersion of bacteria in porous media, our study focuses on the central importance of hydrodynamic flow fluctuations and the active exploration process into high shear regions around the solid grains. The rheotactic coupling between flow and bacteria motility manifests itself at small scales through non-Fickian behavior, and at large scales through a motility-dependent hydrodynamic dispersion effect.
Noticeably, the interplay between fast transport in the flow and motile motion toward grain surfaces is the first necessary step before possible adhesion~\citep{Yates1988}. 
To date, it had been assumed that the transfer between regions of high fluid flow and low flow regions in the vicinity of the grain surfaces was diffusive, like for passive solutes,
and had been modeled as a kinetic single-rate mass transfer process \citep{Yates1988,Bai2016}. Our study suggests that both motility and flow play a central role in the trapping and release processes, which are characterized by two different rates. Both trapping and release rates are proportional to the average flow velocity, while the ratio between mobile and trapped particle increases with increasing flow velocity. The trapping and release mechanisms explain apparently contradictory observations of the concomitant enhancement of retention and dispersion. They are quantified in a theoretical approach that captures the salient features of the  experimental displacement data, and allows for predicting the dispersion of motile bacteria at large scales. These findings shed light on the strategies microorganisms may use to maximize their survival and proliferation abilities under natural conditions, and can give new insights into bacteria filtration and biofilm growth, for which the contact with grain surfaces is determinant. 

\section*{Acknowledgments}
M.D. acknowledges the support of the Spanish Research Agency (10.13039/501100011033) and the Spanish Ministry of Science and Innovation through the project HydroPore (PID2019-106887GB-C31).
H.A and C.D. acknowledge the support of the “Laboratoire d’Excellence Physics Atom Light Mater” (LabEx PALM) as part of the “Investissements d’Avenir” program (reference: ANR-10-LABX-0039). E.C., H..A, C.D. and A.C. are supported by the ANR grant “BacFlow” ANR-15-CE30-0013. EC is supported by the Institut Universitaire de France. Declaration of Interests. The authors report no conflict of interest.

\appendix

\section{Track length statistics\label{app:tracks}}

% In order to study these questions, we used a microfluidics device of
% dimensions xy with porosity x and grain size y and characteristic
% pore length z. We consider different flow rates in
% order to understand the coupling between motility and flow and
% bacteria motion. A strain of so and so bacteria is used both mobile
% and immobile in order to discriminate the difference the impact of
% motility. 

\begin{figure}{h!}
 \centering
\includegraphics[width=0.48\textwidth]{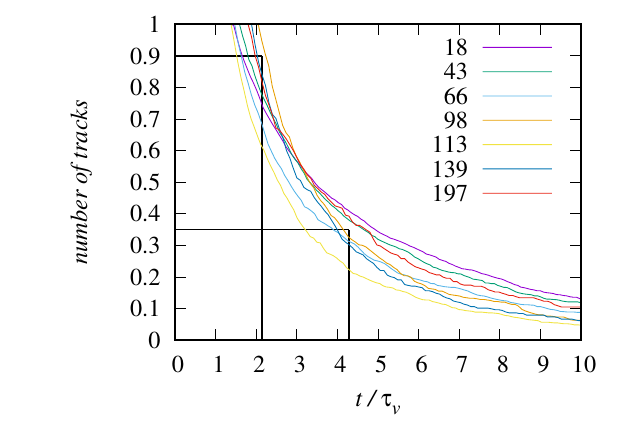}
\includegraphics[width=0.48\textwidth]{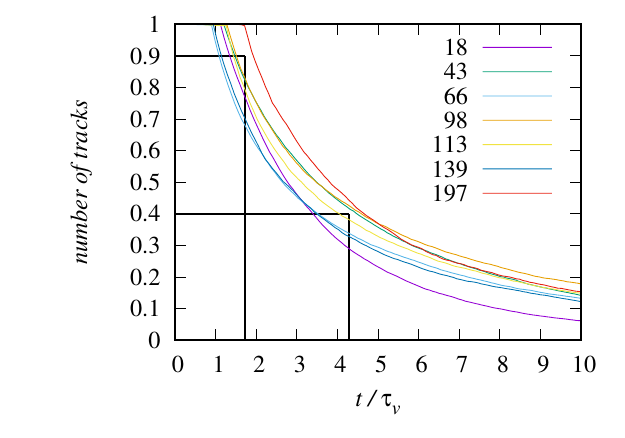}
  \caption{Number of tracks of (left panels) non-motile and (right
    panels) motile bacteria as a function of time. }
  \label{fig:tracks}
\end{figure}

The number of observed tracks decreases with time because tracks leave
the observation window according to their average velocity.
Figure~\ref{fig:tracks} shows the number of tracks of non-motile and
motile bacteria for the experiments at different flow rates as a
function of time measured in units of the characteristic advection time
$\tau_v$, which here is the time to move over the characteristic
grain length $\ell_0$  by mean advection $u_m$. We see that the number
of tracks decreases to around $90\%$ of the initial number of tracks after
around $2 \tau_v$ for the non-motile and motile bacteria.
After around $4 \tau_v$ the number of tracks decreases to
approximately $35\%$ for the non-motile and to around $40\%$ for the motile
bacteria. This means, that the number of tracks of lengths larger than
$4 \ell_0$ is $35\%$ and $40\%$ of the total number of tracks. The
long tracks are tortuous low velocity tracks that can be observed for a
longer time. This is supported by the observation that the mean
velocity starts decreasing after about $2 \tau_v$, as shown in
Figures~\ref{fig:moments:nm} and \ref{fig:moments} below. In the
following, we consider travel times shorter than $5 \tau_v$ in order
to avoid too strong a bias toward slow bacteria. Even so, as we will
see below, there is a significant slowing down of the mean
displacement with increasing travel time, specifically for the
motile bacteria.

%\subsection{Displacement statistics}

\section{Continuous time random walk model\label{app:ctrw}}
\begin{figure}
\includegraphics[width=0.9\textwidth]{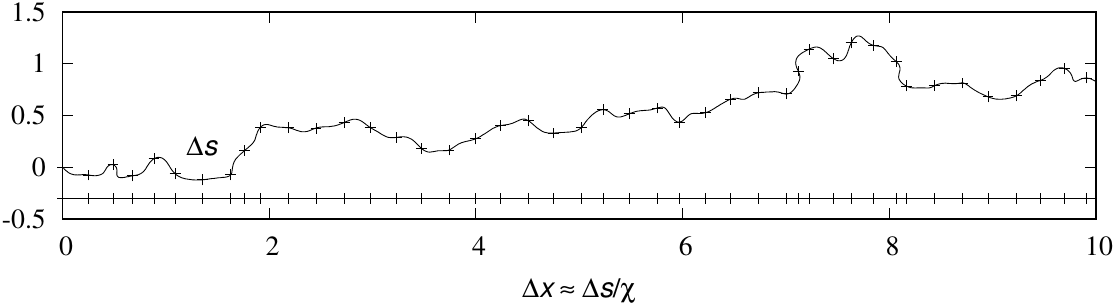}
\caption{Schematic of the representation of a particle trajectory in the CTRW
  approach. Particles speeds are sampled equidistantly at turning points along a
  trajectory indicated by the constant $\Delta s$. The CTRW represents the
  trajectory projected onto the direction of the mean flow.   \label{fig:ctrw}}
\end{figure}

The streamwise motion of a non-motile bacterium in the CTRW is described by
Eq.~\eqref{ctrw:nm}. Unlike classical random walk strategies for the modeling of
particle motion in heterogeneous flow fields, the CTRW approach models particle motion
based on stochastic series of equidistant instead of isochronic particle
speeds~\cite[e.g.,][]{Dentz2016,morales2017}, that is, particle speeds that change
 at equidistant points along a streamline. The streamwise displacement in
the CTRW model represents the projection of the tortuous streamline onto the mean
flow direction using advective tortuosity $\chi$. This is illustrated schematically
in Figure~\ref{fig:ctrw}.
%%%%%%%%%%%%%%%%%%%%%%%%%%%%%%%%%%%%%%%%%%%%%%%
%%%%%%%%%%%%%%%%%%%%%%%%%%%%%%%%%%%%%%%%%%%%%%%
%%%%%%%%%%%%%%%%%%%%%%%%%%%%%%%%%%%%
\subsection{Non-motile bacteria\label{app:ctrwnm}}
%%%%%%%%%%%%%%%%%%%%%%%%%%%%%%%%%%%%
In the following, we provide a derivation of the Boltzmann-type
equation~\eqref{} for the joint distribution $p(x,v,t)$ of particle displacement
and speed. For more details, see~\cite{Comolli2019}. The distribution
$p(x,v,t)$ can be written as
\begin{subequations}
\label{ctrw}
\begin{align}
\label{ctrw:p}
p(x,v,t) = \int\limits_0^t dt' R(x,v,t') \int\limits_{t-t'}^\infty
dt'' \psi(t|v), 
\end{align}
where $\psi(t|v) = \delta(t - \Delta s/v)$. The probability per time
$R(x,v,t)$ for the particle to just arrive at $(x,v)$ at $t$ satisfies
\begin{align}
\label{ctrw:R}
R(x,v,t) = R_0(x,v,t) + \int\limits_0^t dt' \int dx' \int dv'
\psi(x-x', t-t'|v') r(v|v') R(x',v',t'), 
\end{align}
\end{subequations}
where $r(v|v')$ is the transition probability from $v'$ to $v$, and
\begin{align}
\label{psixt}
\psi(x,t|v) = \delta(x - \Delta s/\chi) \delta(t - \Delta s/v).   
\end{align}
The initial condition is encoded in $R_0(x,v,t)$, which is defined by
\begin{align}
R_0(x,v,t) = p_0(x,v) \delta(t),
\end{align}
where $p_0(x,v)$ is the distribution of initial particle positions and
speeds. Equations~\eqref{ctrw:p} and~\eqref{ctrw:R} can be combined in Laplace space to the generalized
master equation
\begin{align}
&\lambda p^\ast(x,v,\lambda) = R_0^\ast(x,v,\lambda) 
\nonumber\\
&+ \int dx' \int
dv' r(v|v') \left[\frac{\lambda \psi^\ast(x-x', \lambda|v')}{1 -
  \psi^\ast(\lambda|v')} p^\ast(x',v',\lambda) - \frac{\lambda
  \psi^\ast(\lambda|v)}{1 - \psi^\ast(\lambda|v)}
p^\ast(x,v,\lambda)\right]. 
\end{align}
Using the explicit form~\eqref{psixt} for $\psi(x,t|v)$, it can be
written as
\begin{align}
&\lambda p^\ast(x,v,\lambda) = R_0^\ast(x,v,\lambda) + \int
dv' r(v|v') \frac{\lambda \exp(-\lambda \Delta s/v')}{1 -
  \exp(-\lambda \Delta s/v')} p^\ast(x - \Delta s/\chi,v',\lambda)
\nonumber\\
&-  \frac{\lambda
  \exp(-\lambda \Delta s/v)}{1 - \exp(-\lambda \Delta s/v)}
p^\ast(x,v,\lambda). 
\end{align}
In the limit of $\Delta s \ll  \ell_c$, we can write
\begin{align}
& \lambda p^\ast(x,v,\lambda) = 
\nonumber\\
&R_0^\ast(x,v,\lambda) + \int
dv' r(v|v') \left[\frac{v'}{\Delta s} p^\ast(x - \Delta s/\chi,v',\lambda) - \frac{v}{\Delta s}
p^\ast(x,v,\lambda)\right]
\\
&
= R_0^\ast(x,v,\lambda) + \int
dv' r(v|v') \frac{v'}{\Delta s} p^\ast(x,v',\lambda) - \frac{v}{\chi}
\frac{\partial}{\partial x} p^\ast(x,v,\lambda) - \frac{v}{\Delta s}
p^\ast(x,v,\lambda),
\end{align}
where we localized $r(v|v') = \delta(v-v')$ in the advection
term. By transformation back to time, we obtain the Boltzmann equation
\begin{align}
\frac{\partial p(x,v,t)}{\partial t} + \frac{v}{\chi} \frac{\partial
  p(x,v,t)}{\partial x} = -\frac{v}{\Delta s} p(x,v,t) +
\int\limits_0^\infty dv' r(v|v') \frac{v'}{\Delta s} p(x,v',t').  
\end{align}
%
%%%%%%%%%%%%%%%%%%%%%%%%%%%%%%
%%%%%%%%%%%%%%%%%%%%%%%%%%%%%%

\subsection{Motile bacteria\label{app:ctrwm}}

In the case of motile bacteria, we account for trapping during
advective steps as well as initial trapping. Thus, we modify~\eqref{ctrw} as
\begin{subequations}
\label{ctrwm}
\begin{align}
\label{ctrwm:p}
p(x,v,t) = R_0(x,v,t) \int\limits_{t}^\infty dt' \psi_0(t|v) + \int\limits_0^t dt' R(x,v,t') \int\limits_{t-t'}^\infty
dt'' \psi_c(t|v).  
\end{align}
We define the initial transition probability
\begin{align}
\label{psi0xt}
\psi_0(x,t|v) &= (1-\rho) \delta(x - \Delta s) \psi_c(t|v) + \rho \delta(x) \psi_f(t|v),
\end{align}
and the distribution of initial transition times
\begin{align}
\psi_0(t|v) &= \int dx \psi_0(x,t|v) = (1-\rho) \psi_c(t|v) + \rho \psi_f(t|v). 
\end{align}
The distribution of compound transition times is given by
\begin{align}
\psi_c(t|v) = \int\limits_0^t dt' \psi(t'|v) \psi_c(t-t'|t'),
\end{align}
where $\psi_c(t-t'|t')$ is defined by~\eqref{psifc}. This relation
reads in Laplace space as
\begin{align}
\psi_c^\ast(\lambda|v) &= \psi^\ast(\lambda[1 - \gamma
\psi_f^\ast(\lambda)]) = \exp(-\lambda[1 - \gamma
\psi_f^\ast(\lambda)] \Delta s/v).
\end{align}
The probability per time $R(x,v,t)$ for the particle to just arrive at $(x,v)$ at $t$ satisfies
\begin{align}
\label{ctrwm:R}
R(x,v,t) = R_1(x,v,t) + \int\limits_0^t dt' \int dx' \int dv'
\psi_0(x-x', t-t'|v') r(v|v') R(x',v',t'), 
\end{align}
\end{subequations}
where $r(v|v')$ is the transition probability from $v'$ to $v$, and
\begin{align}
\label{psimxt}
\psi_c(x,t|v) = \delta(x - \Delta s) \psi_c(t|v).   
\end{align}
The $R_1(x,v,t)$ is given by
\begin{align}
\label{R1}
R_1(x,v,t) = \int\limits_0^t dt' \int dx' \int dv'
\psi_0 (x-x', t-t'|v') r(v|v') R_0(x',v',t').  
\end{align}

Equations~\eqref{ctrwm:p} and~\eqref{ctrwm:R} can be combined in Laplace space to the generalized
master equation
\begin{align}
\label{gmem}
\lambda G^\ast(x,v,\lambda) &= R_1^\ast(x,v,\lambda) + 
\int dv' r(v|v') \frac{\lambda
  \psi_c^\ast(\lambda|v')}{1-\psi_c^\ast(\lambda|v')}  G^\ast(x -
\Delta s/\chi,v',\lambda)
\nonumber\\
& - \frac{\lambda
  \psi_c^\ast(\lambda|v)}{1-\psi_c^\ast(\lambda|v)} G^\ast(x,v,\lambda),
\end{align}
where we defined
\begin{align}
G^\ast(x,v,\lambda) = \left[p^\ast(x,v,\lambda) -
R_0^\ast(x,v,\lambda) \frac{1 - \psi_0(\lambda|v)}{\lambda} \right]. 
\end{align}
Using this definition and definition~\eqref{R1}, we can write~\eqref{gmem} as
\begin{align}
&\lambda p^\ast(x,v,\lambda) =  R_0^\ast(x,v,\lambda)  
\nonumber\\
& +
\int dx' \int dv' r(v|v') \left[ \psi_0^\ast (x-x',\lambda|v')
R_0^\ast(x',v',\lambda) - \psi_0^\ast(\lambda|v) R_0^\ast(x,v,\lambda) \right]
\nonumber\\
& + \int dv' r(v|v') \left[\frac{\lambda
  \psi_c^\ast(\lambda|v')}{1-\psi_c^\ast(\lambda|v')}  G^\ast(x -
\Delta s/\chi,v',\lambda) - \frac{\lambda
  \psi_c^\ast(\lambda|v)}{1-\psi_c^\ast(\lambda|v)} G^\ast(x,v,\lambda)\right]. 
\end{align}
Using the definition~\eqref{psi0xt} of $\psi_0(x,t|v)$, we obtain
\begin{align}
&\lambda p^\ast(x,v,\lambda) =  R_0^\ast(x,v,\lambda)  
\nonumber\\
& +
\int dx' \int dv' r(v|v') (1 - \rho) \left[\psi_c^\ast (\lambda|v')
R_0^\ast(x - \Delta s/\chi,v',\lambda) - \psi_c^\ast(\lambda|v) R_0^\ast(x,v,\lambda) \right]
\nonumber\\
& + \int dv' r(v|v') \left[\frac{\lambda
  \psi_c^\ast(\lambda|v')}{1-\psi_c^\ast(\lambda|v')}  G^\ast(x -
\Delta s/\chi,v',\lambda) - \frac{\lambda
  \psi_c^\ast(\lambda|v)}{1-\psi_c^\ast(\lambda|v)} G^\ast(x,v,\lambda)\right]. 
\end{align}
Note that 
\begin{align}
&\frac{\lambda \psi_c^\ast(\lambda|v)}{1-\psi_c^\ast(\lambda|v)}
R_0^\ast(x,v,\lambda) \frac{1 - \psi_0(\lambda|v)}{\lambda}
\nonumber
\\
&= R_0^\ast(x,v,\lambda) (1 - \rho) \psi_c^\ast(\lambda|v) +
\rho R_0^\ast(x,v,\lambda)
\frac{\psi_c^\ast(\lambda|v)}{1-\psi_c^\ast(\lambda|v)} \phi^\ast(\lambda), 
\end{align}
where we defined
\begin{align}
\label{app:phi}
\phi^\ast(\lambda) = \frac{1 - \psi^\ast_f(\lambda)}{\lambda}. 
\end{align}
Combining everything, we obtain 
\begin{align}
\lambda p^\ast(x,v,\lambda) &=  R_0^\ast(x,v,\lambda)  + \int dv' r(v|v') \frac{\lambda
  \psi_c^\ast(\lambda|v')}{1-\psi_c^\ast(\lambda|v')}  G_m^\ast(x -
\Delta s/\chi,v',\lambda) 
\nonumber\\
& - \frac{\lambda
  \psi_c^\ast(\lambda|v)}{1-\psi_c^\ast(\lambda|v)} G_m^\ast(x,v,\lambda),
\end{align}
where we defined
\begin{align}
G^\ast_m(x,v,\lambda) = p^\ast(x,v,\lambda) - \rho
R_0^\ast(x,v,\lambda) \phi^\ast(\lambda). 
\end{align}
Furthermore, we approximate for small $\Delta s$
\begin{align}
\frac{\psi_c^\ast(\lambda|v)}{1-\psi_c^\ast(\lambda|v)} = \frac{v}{\Delta
s} \frac{1}{1+\gamma \phi^\ast(\lambda)}  
\end{align}
Thus, we obtain 
\begin{align}
\lambda p^\ast(x,v,\lambda) &=  R_0^\ast(x,v,\lambda)  + \int dv'
r(v|v') \frac{v'}{\Delta s}  \frac{G_m^\ast(x -
\Delta s/\chi,v',\lambda)}{1+\gamma \phi^\ast(\lambda)}  
\nonumber\\
&- \frac{v}{\Delta s} \frac{G_m^\ast(x,v,\lambda)}{1+\gamma \phi^\ast(\lambda)},
\end{align}
We define now the mobile concentration of bacteria in the stream as
\begin{align}
p_s^\ast(x,v,\lambda) = \frac{G_m^\ast(x,v,\lambda)}{1+\gamma
  \phi^\ast(\lambda)} = \frac{p^\ast(x,v,\lambda) - \rho
R_0^\ast(x,v,\lambda) \phi^\ast(\lambda)}{1+\gamma \phi^\ast(\lambda)}
\end{align}
With this definition, we obtain
\begin{align}
\lambda [1 + \gamma \phi^\ast(\lambda)]& p_s^\ast(x,v,\lambda) =
R_0^\ast(x,v,\lambda) - \rho \lambda
R_0^\ast(x,v,\lambda) \phi^\ast(\lambda)
\nonumber\\
&+ \int dv'
r(v|v') \left[\frac{v'}{\Delta s}  p_s^\ast(x -
\Delta s/\chi,v',\lambda)  - \frac{v}{\Delta s} p_s^\ast(x,v,\lambda) \right],
\end{align}
Expanding the integral terms on the right side in analogy to the
previous section gives 
\begin{align}
\lambda [1 + \gamma \phi^\ast(\lambda)]& p_s^\ast(x,v,\lambda) =
R_0^\ast(x,v,\lambda) - \rho \lambda
R_0^\ast(x,v,\lambda) \phi^\ast(\lambda)
\nonumber\\
& + \int
dv' r(v|v') \frac{v'}{\Delta s} p^\ast(x,v',\lambda) - \frac{v}{\chi}
\frac{\partial}{\partial x} p^\ast(x,v,\lambda) - \frac{v}{\Delta s}
p^\ast(x,v,\lambda),
\end{align}
By transformation back to time, we obtain the Boltzmann equation
\begin{align}
\frac{\partial p_s(x,v,t)}{\partial t} &+ \frac{\partial}{\partial t}
\int\limits_0^t dt' \gamma \phi(t - t') p_s(x,t') +
\frac{v}{\chi} \frac{\partial
  p(x,v,t)}{\partial x} =
\nonumber\\
&  \rho p_0(x,v) \psi_f(t) -\frac{v}{\Delta s} p(x,v,t) +
\int\limits_0^\infty dv' r(v|v') \frac{v'}{\Delta s} p(x,v',t').  
\end{align}

\section{Initial trapping time distribution\label{app:psi0}}

In order to derive the initial trapping time distribution, we employ the concept of the backward recurrence time $B_{t_0} = t_0 - t_N$, this means the time that has passed between a target time $t_0$ and the time $t_N$ of the last trapping event before $t_0$. For a Poissonian trapping process, this means for an exponential inter-event time distribution, the distribution of $B_{t_0}$ in the steady state limit, that is, for $N \to \infty$, is given by~\cite[][]{Godreche2001}
\begin{align}
\label{psiB}
\psi_B(t) = \gamma \exp(-\gamma t). 
\end{align}
It is independent from $t_0$, $B_{t_0} \equiv B$. The initial trapping time $\eta_0$ can be expressed in terms of $B$ as $\eta_0 = \tau_f - B$. 
Thus, the joint distribution for a bacterium to be trapped and have the trapping time $\eta_0 < t$ is 
\begin{align}
\text{Prob}(\eta_0 < t  \land \text{trapped}) = \left\langle H(\tau_f - B) H[t - (\tau_f - B)]  \right\rangle 
\end{align}
It can be written as
\begin{align}
\text{Prob}(\eta_0 < t  \land \text{trapped}) = \int\limits_0^\infty dt' \int\limits_{0}^\infty d t''  H(t' - t'') H[t - (t' - t'')] \psi_B(t'') \psi_f(t').
\end{align}

Using expression~\eqref{psiB} for $\psi_B(t)$ and shifting $t''
\to t' -t''$, we obtain
\begin{align}
& \text{Prob}(\eta_0 < t  \land \text{trapped}) = \int\limits_0^\infty
dt' \int\limits_{0}^{\infty} d t'' \gamma \exp[-\gamma (t'-t'')]
\psi_f(t') H(t - t'') H(t'-t'')
\nonumber\\
&= \int\limits_0^t
dt'' \int\limits_{t''}^{\infty} d t' \gamma \exp[-\gamma (t'-t'')]
\psi_f(t')
\label{psi0psif}
\end{align}
Thus we obtain for the joint probabiility of being trapped and
$\eta_0$ in $[t,t+dt]$ by derivation of~\eqref{psi0psif} with respect
to $t$
\begin{align}
\text{P}_0(t) = \int\limits_{t}^{\infty} d t' \gamma \exp[-\gamma (t'-t)] \psi_f(t')
\end{align}
For $\psi_f(t) = \exp(-t/\tau_c)/\tau_c$, we obtain 
\begin{align}
& \text{Prob}(\eta_0 < t  \land \text{trapped}) = \int\limits_0^t dt''
\int\limits_{t''}^{\infty} d t' \gamma \tau_c^{-1} \exp[\gamma t'' - t'
(\tau_c^{-1} + \gamma)]
\nonumber\\
&= \frac{\gamma \tau_c}{1 + \gamma \tau_c} \exp(-t/\tau_c).  
\end{align}
%
%%%%%%%%%%%%%%%%%%%%%%%%%%%%%%%%%%%%%%%
\section{Asymptotic dispersion and retardation coefficients for motile bacteria\label{app:asymptotic}}
%%%%%%%%%%%%%%%%%%%%%%%%%%%%%%%%%%%%%%%
In order to derive the dispersion and retardation coefficients for
motile bacteria, we consider the Fourier-Laplace transform of the
total bacteria distribution $p(x,t) = p_s(x,t) + p_g(x,t)$. From the
Fourier-Laplace transform of \eqref{pgmrmt}, we obtain
\begin{align}
\label{app:pk}
\tilde{p}^\ast(k,\lambda) = \tilde{p}_s^\ast(k,\lambda)\left[1 +
\phi^\ast(\lambda) \gamma\right] + \rho \phi^\ast(\lambda).
\end{align}
The Fourier-Laplace transform of the density $p_s(x,t)$ in the stream
is obtained from~\eqref{eq:mrmt} as
\begin{align}
\label{app:psk}
\tilde{p}_s^\ast(k,\lambda) = \frac{1 - \rho \lambda \phi^\ast(\lambda)}{\lambda \left[1 + \phi^\ast(\lambda) \gamma\right] - i k u_m + D_{nm} k^2},
\end{align}
where we used Eq.~\eqref{app:phi} to express $\psi^\ast_f(\lambda)$ in
terms of $\phi^\ast(\lambda)$. The Laplace transforms of the mean and mean square displacements are given in terms of $\tilde{p}^\ast(k,\lambda)$ as 
\begin{align}
m_n^\ast(\lambda) &= (- i)^n \left.\frac{\partial^n \tilde{p}^\ast(k,\lambda)}{\partial k^n}\right|_{k = 0} 
\end{align}
for $n = 1, 2$. Using~\eqref{app:pk}, we obtain
\begin{align}
m_n^\ast(\lambda) &= (- i)^n \left.\frac{\partial^n \tilde{p}_s^\ast(k,\lambda)}{\partial k^n}\right|_{k = 0} \left[1 + \phi^\ast(\lambda) \gamma\right].    
\end{align}
Using~\eqref{app:psk}, we obtain the explicit expressions
\begin{align}
m_1^\ast(\lambda) &=  \frac{u_m}{\lambda^2}\frac{1 - \rho \lambda \phi^\ast(\lambda)}{\left[1 + \phi^\ast(\lambda) \gamma\right]} 
\\
m_2^\ast(\lambda) &=  \frac{2 D}{\lambda^2}\frac{1 - \rho \lambda \phi^\ast(\lambda)}{\left[1 + \phi^\ast(\lambda) \gamma\right]} + \frac{2 u_m^2}{\lambda^3} \frac{1 - \rho \lambda \phi^\ast(\lambda)}{\left[1 + \phi^\ast(\lambda) \gamma\right]^2}
\end{align}

We set now $\rho = \beta/(1 + \beta)$ with $\beta = \gamma \tau_c$ and $\phi(t) = \exp(-t/\tau_c)$, which implies
\begin{align}
\phi^\ast(\lambda) = \frac{\tau_c}{1 + \lambda \tau_c}. 
\end{align}
Thus, we obtain 
\begin{align}
m_1^\ast(\lambda) &=  \frac{u_m}{(1 + \beta)\lambda^2}
\\
m_2^\ast(\lambda) &=  \frac{2 D_{nm}}{(1 + \beta)\lambda^2} + \frac{2 u_m^2}{(1 + \beta)^2 \lambda^3} \frac{1+\lambda \tau_c}{1+\lambda \frac{\tau_c}{1 + \beta}} 
\end{align}
The latter can be written as 
\begin{align}
  m_2^\ast(\lambda) &=  \frac{2 D_{nm}}{(1 + \beta)\lambda^2} + \frac{2 u_m^2}{(1 + \beta)^2 \lambda^3} +  
\frac{2 u_m^2 \beta \tau_c}{(1 + \beta)^3 \lambda^2} \frac{1}{1+\lambda \frac{\tau_c}{1 + \beta}} 
\end{align}
In the limit of $\lambda \tau_c \to 0$, we obtain in leading order 
\begin{align}
 m_2^\ast(\lambda) &=  \frac{2 D_{nm}}{(1 + \beta)\lambda^2} + \frac{2 u_m^2}{(1 + \beta)^2 \lambda^3} +  
\frac{2 u_m^2 \beta \tau_c}{(1 + \beta)^3 \lambda^2} 
\end{align}
Inverse Laplace transform gives 
\begin{align}
m_1(t) &= \frac{u_m t}{1 + \beta}
\\
m_2(t) &= \frac{2D_{nm} t}{1 + \beta} + \frac{u_m^2 t^2}{(1 + \beta)^2} + \frac{2 u_m^2 \beta \tau_c}{(1 + \beta)^3}
\end{align}
We define the retardation coefficient by comparing $m_1(t)$ with the mean displacement for the non-motile bacteria. This gives
\begin{align}
\label{app:R}
R = 1 + \beta. 
\end{align}
The displacement variance is given by
\begin{align}
\sigma^2(t) = \frac{2 D_{nm} t}{R} + \frac{2 u_m^2 \tau_c (R-1) t}{R^3}
\end{align}
Thus, we obtain for the dispersion coefficient 
\begin{align}
\label{app:Dm}
D_m = \frac{D_{nm}}{R} + \frac{u_m^2 \tau_c (R-1)}{R^3}
\end{align}

We consider now the asymptotic equation for the total bacteria
concentration. 
% In order to derive the dispersion and retardation coefficients for
% motile bacteria,
Thus, we consider the Fourier-Laplace transform of the
total bacteria distribution $p(x,t) = p_s(x,t) + p_g(x,t)$. From the
Fourier-Laplace transform of \eqref{pgmrmt}, we obtain
\begin{align}
%\tilde{p}^\ast(k,\lambda) = \tilde{p}_s^\ast(k,\lambda)\left[1 +
%\phi^\ast(\lambda) \gamma\right] + \rho \phi^\ast(\lambda).
\tilde{p}_s^\ast(k,\lambda) = \frac{\tilde{p}^\ast(k,\lambda) - \rho \phi^\ast(\lambda)}{\left[1 + \phi^\ast(\lambda) \gamma\right]}.  
\end{align}
Thus, we obtain from~\eqref{eq:mrmt}
\begin{align}
\label{app:c}
\lambda \tilde{p}^\ast(k,\lambda) - \left( i k u_m - D_{nm} k^2\right)
\frac{\tilde{p}^\ast(k,\lambda) - \rho \phi^\ast(\lambda) }{1 +
  \phi^\ast(\lambda) \gamma} = 1, 
\end{align}
where we used Eq.~\eqref{app:phi} to express $\psi^\ast_f(\lambda)$ in
terms of $\phi^\ast(\lambda)$. We use the expansion 
\begin{align}
\phi^\ast(\lambda) = \tau_c \left(1 - \lambda \tau_c \right). 
\end{align}
in order to expand~\eqref{app:c} up to linear order in $\lambda$
\begin{align}
\lambda \tilde{p}^\ast(k,\lambda) - \frac{i k u_m - D_{nm}
  k^2}{1 + \gamma \tau_c} \tilde{p}^\ast(k,\lambda)
\left(1 - \frac{\lambda \gamma \tau_c^2}{1 + \gamma \tau_c} \right)= 1,
\end{align}
where we disregard terms of order $k \phi^\ast(\lambda)$ and order
$\lambda^2$. We set now self-consistently
\begin{align}
\lambda \tilde{p}^\ast(k,\lambda) = 1 + \frac{i k u_m}{1 + \gamma
  \tau_c} \tilde{p}^\ast(k,\lambda)
\end{align}
to obtain
\begin{align}
\lambda \tilde{p}^\ast(k,\lambda) - \frac{i k u_m - D_{nm}
  k^2}{1 + \gamma \tau_c} \tilde{p}^\ast(k,\lambda) +
\frac{u_m^2 \gamma \tau_c^2 k^2}{(1 + \gamma \tau_c)^3} \tilde{p}^\ast(k,\lambda)= 1,
\end{align}
where we disregard terms of order $k$. Using definitions~\eqref{app:R}
and~\eqref{app:Dm}, we obtain 
\begin{align}
\lambda \tilde{p}^\ast(k,\lambda) - \left(i k \frac{u_m}{R} - D_{m}
k^2 \right) \tilde{p}^\ast(k,\lambda)= 1. 
\end{align}
The inverse Fourier-Laplace transform of this equation gives
Eq.~\eqref{ADE:pm}. 

\section{Physical model for bacteria blow-off from
  grains \label{app:arrachement}}
A simple model is proposed with the objective of
showing that the characteristic residence time $\tau_c$ is inversely proportional to the average flow
velocity. Let's consider a circular obstacle of size $l_0$ facing a
flow of average velocity $U$. The flow field around the grain is given
by
\begin{align}
v_r &=  U \left(1- \frac{\ell_0^2}{4r^2} \right) \cos(\theta)
\\
  v_{\theta} &= - U \left(1+ \frac{\ell_0^2}{4 r^2}\right) \sin(\theta),
\end{align}
where $r$ is the distance from the center of the grain and $\theta$
the angle with respect to the flow direction. The shear rate on the grain surface is:
\begin{align}
\dot \gamma = \left.\frac{\partial v_{\theta}}{\partial
  r}\right|_{r=\frac{\ell_0}{2}}=\frac{4U}{\ell_0} \sin(\theta)
\end{align}
Bacteria transported in the vicinity of the grain rotate because of
the local shear. Because of their swimming ability, some are able
to reach the rear of the obstacles where the flow is
low~\citep{Mino2018}. Once on the surface, the bacteria body aligns
with the surface and hydrodynamic interaction favors their swimming
along the surface. Hydrodynamic interactions are know to influence the
bacteria over a distance $\delta$ of the order of ten
microns~\citep{Berke2008,Li2011}. As they move upstream along the
surface, they face an increasing shear rate. When the shear rate
reaches the critical value of $\dot\gamma_c \sim 5 s^{-1}$, the bacteria are stopped by the
flow and are eventually detached from the surface and returned to the
flow. This scenario is based on the video available in the supplemental material
section of~\cite{Creppy2019}. This video shows
motile bacteria (white rods) transported by a flow (average velocity
72 $\mu m/s$) . On the video, the upstream displacements are clearly
identifiable as well as the motion towards the rear of the grains and
the displacements on the surfaces and the final release. This
succession of steps was also recently identified by computer
simulations using molecular dynamics coupled with lattice
Boltzmann~\citep{Lee2021} as the scenario characterizing the entrapment
and release of motile bacteria moving near an obstacle.

The critical shear rate is reached when
$\theta=\arcsin(\dot\gamma_c \ell_0/4U)$. The model requires a
minimal mean flow velocity $U_c = \frac{\ell_0 \dot\gamma_c}{4}$, below
which diffusion of the bacteria due to the swimming activity
dominates. The minimal fluid velocity required to see the separation
between bacteria moving on the grains and in the pore channels
 is about $30 \mu$m/s. Above this velocity,  the total
distance swum by the bacteria on the grain surface before its release
is $l\sim \ell_0 \theta/2$ if $\theta$ is not too large.
The motion on the grain is at swimming velocity $v_0$ and the total
time to swim from the back of the grain to the critical angle is
$\tau_c = \frac{\ell_0^2 \dot\gamma_c}{8 v_0} \frac{1}{U}$. We recover here
the scaling obtained from interpretation of the data by the CTRW
model.  

\begin{figure}

\includegraphics[width=0.45\textwidth]{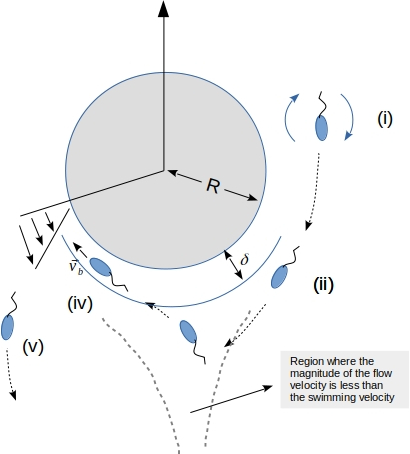}

\caption{Illustration of the model. (i) because of the local shear, the bacteria rotates in the flow (ii) some are redirected towards the rear of the grain where the flow velocity is small (iii) the bacteria swims towards the grain and then along the surface. As they move along the grain they face an increase local shear rate. When the shear become larger than a critical value$\dot{\gamma}_c$, the bacteria gets blow and goes back to the flow.} 
\label{fig:model}
\end{figure}
%%%%%%%%%%%%%%%%%%%%%%%%%%%%%%%%%%%%%%%%%%%%
\bibliographystyle{jfm}
\bibliography{bacteria}

\begin{thebibliography}{66}
\expandafter\ifx\csname natexlab\endcsname\relax\def\natexlab#1{#1}\fi
\def\au#1{#1} \def\ed#1{#1} \def\yr#1{#1}\def\at#1{#1}\def\jt#1{\textit{#1}}
  \def\bt#1{#1}\def\bvol#1{\textbf{#1}} \def\vol#1{#1} \def\pg#1{#1}
  \def\publ#1{#1}\def\arxiv#1{#1}\def\org#1{#1}\def\st#1{\textit{#1}}

\bibitem[Alim {\em et~al.\/}(2017)Alim, Parsa, Weitz \& Brenner]{alim2017}
{\sc \au{Alim, Karen}, \au{Parsa, Shima}, \au{Weitz, David~A.} \& \au{Brenner,
  Michael~P.}} \yr{2017}  \at{Local pore size correlations determine flow
  distributions in porous media}.  \jt{Physical Review Letters}
  \bvol{119}~(14).

\bibitem[Alonso-Matilla {\em et~al.\/}(2019)Alonso-Matilla, Chakrabarti \&
  Saintillan]{Alonso2019}
{\sc \au{Alonso-Matilla, Roberto}, \au{Chakrabarti, Brato} \& \au{Saintillan,
  David}} \yr{2019}  \at{Transport and dispersion of active particles in
  periodic porous media}.  \jt{Phys. Rev. Fluids}  \bvol{4},  \pg{043101}.

\bibitem[Altshuler {\em et~al.\/}(2013)Altshuler, Mi{\~n}o, P{\'e}rez-Penichet,
  R{\'\i}o, Lindner, Rousselet \& Cl{\'e}ment]{Altshuler2013}
{\sc \au{Altshuler, E.}, \au{Mi{\~n}o, G.}, \au{P{\'e}rez-Penichet, C.},
  \au{R{\'\i}o, L.~del}, \au{Lindner, A.}, \au{Rousselet, A.} \&
  \au{Cl{\'e}ment, E.}} \yr{2013}  \at{Flow-controlled densification and
  anomalous dispersion of e. coli through a constriction}.  \jt{Soft Matter}
  \bvol{9}~(6),  \pg{1864--1870}.

\bibitem[de~Anna {\em et~al.\/}(2020)de~Anna, Pahlavan, Yawata, Stocker \&
  Juanes]{deAnna2020}
{\sc \au{de~Anna, Pietro}, \au{Pahlavan, Amir~A.}, \au{Yawata, Yutaka},
  \au{Stocker, Roman} \& \au{Juanes, Ruben}} \yr{2020}  \at{Chemotaxis under
  flow disorder shapes microbial dispersion in porous media}.  \jt{Nature
  Physics}  \bvol{17}~(1),  \pg{68--73}.

\bibitem[Aramideh {\em et~al.\/}(2018)Aramideh, Vlachos \&
  Ardekani]{Aramideh2018}
{\sc \au{Aramideh, Soroush}, \au{Vlachos, Pavlos~P.} \& \au{Ardekani,
  Arezoo~M.}} \yr{2018}  \at{Pore-scale statistics of flow and transport
  through porous media}.  \jt{Physical Review E}  \bvol{98}~(1).

\bibitem[Bai {\em et~al.\/}(2016)Bai, Cochet, Pauss \& Lamy]{Bai2016}
{\sc \au{Bai, Hongjuan}, \au{Cochet, Nelly}, \au{Pauss, Andr{\'e}} \& \au{Lamy,
  Edvina}} \yr{2016}  \at{Bacteria cell properties and grain size impact on
  bacteria transport and deposition in porous media}.  \jt{Colloids and
  Surfaces B: Biointerfaces}  \bvol{139},  \pg{148--155}.

\bibitem[Bear(1972)]{Bear:1972}
{\sc \au{Bear, J.}} \yr{1972} {\em Dynamics of fluids in porous media\/}.
  \publ{American Elsevier, New York}.

\bibitem[Becker {\em et~al.\/}(2003)Becker, Metge, Collins, Shapiro \&
  Harvey]{Becker2003}
{\sc \au{Becker, Matthew~W.}, \au{Metge, David~W.}, \au{Collins, Samantha~A.},
  \au{Shapiro, Allen~M.} \& \au{Harvey, Ronald~W.}} \yr{2003}  \at{Bacterial
  transport experiments in fractured crystalline bedrock}.  \jt{Ground Water}
  \bvol{41}~(5),  \pg{682--689}.

\bibitem[Berg(2018)]{berg2018random}
{\sc \au{Berg, Howard~C}} \yr{2018}  \at{Random walks in biology}.  \bt{In {\em
  Random Walks in Biology\/}}.  \publ{Princeton University Press}.

\bibitem[Berke {\em et~al.\/}(2008)Berke, Turner, Berg \& Lauga]{Berke2008}
{\sc \au{Berke, Allison~P.}, \au{Turner, Linda}, \au{Berg, Howard~C.} \&
  \au{Lauga, Eric}} \yr{2008}  \at{Hydrodynamic attraction of swimming
  microorganisms by surfaces}  \bvol{101}~(3).

\bibitem[Berkowitz \& Scher(1997)]{BS1997}
{\sc \au{Berkowitz, B.} \& \au{Scher, H.}} \yr{1997}  \at{Anomalous transport
  in random fracture networks}.  \jt{Phys. Rev. Lett.}  \bvol{79}~(20),
  \pg{4038--4041}.

\bibitem[Brenner \& Edwards(1993)]{brenner}
{\sc \au{Brenner, H.} \& \au{Edwards, D.}} \yr{1993} {\em Macrotransport
  Processes\/}.  \publ{Butterworth-Heinemann, MA, USA}.

\bibitem[Camesano \& Logan(1998)]{Camesano1998}
{\sc \au{Camesano, Terri~A.} \& \au{Logan, Bruce~E.}} \yr{1998}  \at{Influence
  of fluid velocity and cell concentration on the transport of motile and
  nonmotile bacteria in porous media}.  \jt{Environmental Science {\&}
  Technology}  \bvol{32}~(11),  \pg{1699--1708}.

\bibitem[Carrel {\em et~al.\/}(2018)Carrel, Morales, Dentz, Derlon, Morgenroth
  \& Holzner]{carrel2018}
{\sc \au{Carrel, Maxence}, \au{Morales, Ver{\'o}nica~L}, \au{Dentz, Marco},
  \au{Derlon, Nicolas}, \au{Morgenroth, Eberhard} \& \au{Holzner, Markus}}
  \yr{2018}  \at{Pore-scale hydrodynamics in a progressively bioclogged
  three-dimensional porous medium: 3-d particle tracking experiments and
  stochastic transport modeling}.  \jt{Water resources research}
  \bvol{54}~(3),  \pg{2183--2198}.

\bibitem[Comolli {\em et~al.\/}(2019)Comolli, Hakoun \& Dentz]{Comolli2019}
{\sc \au{Comolli, Alessandro}, \au{Hakoun, Vivien} \& \au{Dentz, Marco}}
  \yr{2019}  \at{Mechanisms, upscaling, and prediction of anomalous dispersion
  in heterogeneous porous media}.  \jt{Water Resources Research}
  \bvol{55}~(10),  \pg{8197--8222}.

\bibitem[Creppy {\em et~al.\/}(2019)Creppy, Cl\'ement, Douarche, D'Angelo \&
  Auradou]{Creppy2019}
{\sc \au{Creppy, Adama}, \au{Cl\'ement, Eric}, \au{Douarche, Carine},
  \au{D'Angelo, Maria~Veronica} \& \au{Auradou, Harold}} \yr{2019}  \at{Effect
  of motility on the transport of bacteria populations through a porous
  medium}.  \jt{Phys. Rev. Fluids}  \bvol{4},  \pg{013102}.

\bibitem[Darcy(1856)]{darcy1856fontaines}
{\sc \au{Darcy, Henry}} \yr{1856} {\em Les fontaines publiques de la ville de
  Dijon: exposition et application...\/}.  \publ{Victor Dalmont}.

\bibitem[De~Anna {\em et~al.\/}(2017)De~Anna, Quaife, Biros \&
  Juanes]{deannaprf2017}
{\sc \au{De~Anna, P.}, \au{Quaife, B.}, \au{Biros, G.} \& \au{Juanes, R.}}
  \yr{2017}  \at{Prediction of velocity distribution from pore structure in
  simple porous media}.  \jt{Phys. Rev. Fluids}  \bvol{2},  \pg{124103}.

\bibitem[Dehkharghani {\em et~al.\/}(2019)Dehkharghani, Waisbord, Dunkel \&
  Guasto]{Dehkharghani2019}
{\sc \au{Dehkharghani, Amin}, \au{Waisbord, Nicolas}, \au{Dunkel, J\"{o}rn} \&
  \au{Guasto, Jeffrey~S.}} \yr{2019}  \at{Bacterial scattering in microfluidic
  crystal flows reveals giant active taylor{\textendash}aris dispersion}.
  \jt{Proceedings of the National Academy of Sciences}  \bvol{116}~(23),
  \pg{11119--11124}.

\bibitem[Dentz \& Berkowitz(2003)]{dentz2003transport}
{\sc \au{Dentz, Marco} \& \au{Berkowitz, Brian}} \yr{2003}  \at{Transport
  behavior of a passive solute in continuous time random walks and multirate
  mass transfer}.  \jt{Water Resources Research}  \bvol{39}~(5).

\bibitem[Dentz {\em et~al.\/}(2018)Dentz, Icardi \& Hidalgo]{Dentz2018jfm}
{\sc \au{Dentz, M.}, \au{Icardi, M.} \& \au{Hidalgo, J.~J.}} \yr{2018}
  \at{Mechanisms of dispersion in a porous medium}.  \jt{J. Fluid Mech.}
  \bvol{841},  \pg{851--882}.

\bibitem[Dentz {\em et~al.\/}(2016)Dentz, Kang, Comolli, Le~Borgne \&
  Lester]{Dentz2016}
{\sc \au{Dentz, Marco}, \au{Kang, Peter~K}, \au{Comolli, Alessandro},
  \au{Le~Borgne, Tanguy} \& \au{Lester, Daniel~R}} \yr{2016}  \at{Continuous
  time random walks for the evolution of lagrangian velocities}.  \jt{Physical
  Review Fluids}  \bvol{1}~(7),  \pg{074004}.

\bibitem[Feller(1968)]{Feller:1}
{\sc \au{Feller, W.}} \yr{1968} {\em An Introduction to Probability Theory and
  Its Applications\/},  \st{Wiley Series in Probability and Statistics},
  \vol{vol.~1}.  \publ{Wiley}.

\bibitem[Figueroa-Morales {\em et~al.\/}(2015)Figueroa-Morales,
  Leonardo~Mi{\~n}o, Rivera, Caballero, Cl{\'e}ment, Altshuler \&
  Lindner]{Figueroa2015}
{\sc \au{Figueroa-Morales, Nuris}, \au{Leonardo~Mi{\~n}o, Gast{\'o}n},
  \au{Rivera, Aramis}, \au{Caballero, Rogelio}, \au{Cl{\'e}ment, Eric},
  \au{Altshuler, Ernesto} \& \au{Lindner, Anke}} \yr{2015}  \at{Living on the
  edge: transfer and traffic of e. coli in a confined flow}.  \jt{Soft Matter}
  \bvol{11}~(31),  \pg{6284--6293}.

\bibitem[Figueroa-Morales {\em et~al.\/}(2020{\natexlab{{\em
  a\/}}})Figueroa-Morales, Rivera, Soto, Lindner, Altshuler \&
  Cl{\'{e}}ment]{Figueroa-Morales2020}
{\sc \au{Figueroa-Morales, Nuris}, \au{Rivera, Aramis}, \au{Soto, Rodrigo},
  \au{Lindner, Anke}, \au{Altshuler, Ernesto} \& \au{Cl{\'{e}}ment,
  {\'{E}}ric}} \yr{2020{\natexlab{{\em a\/}}}}  \at{E. coli
  {\textquotedblleft}super-contaminates{\textquotedblright} narrow ducts
  fostered by broad run-time distribution}.  \jt{Science Advances}
  \bvol{6}~(11).

\bibitem[Figueroa-Morales {\em et~al.\/}(2020{\natexlab{{\em
  b\/}}})Figueroa-Morales, Soto, Junot, Darnige, Douarche, Martinez, Lindner \&
  Cl{\'e}ment]{figueroa20203d}
{\sc \au{Figueroa-Morales, Nuris}, \au{Soto, Rodrigo}, \au{Junot, Gaspard},
  \au{Darnige, Thierry}, \au{Douarche, Carine}, \au{Martinez, Vincent~A},
  \au{Lindner, Anke} \& \au{Cl{\'e}ment, Eric}} \yr{2020{\natexlab{{\em b\/}}}}
   \at{3d spatial exploration by e. coli echoes motor temporal variability}.
  \jt{Physical Review X}  \bvol{10}~(2),  \pg{021004}.

\bibitem[Ghanbarian {\em et~al.\/}(2013)Ghanbarian, Hunt, Ewing \&
  Sahimi]{Ghanbarian2013}
{\sc \au{Ghanbarian, Behzad}, \au{Hunt, Allen~G.}, \au{Ewing, Robert~P.} \&
  \au{Sahimi, Muhammad}} \yr{2013}  \at{Tortuosity in porous media: A critical
  review}.  \jt{Soil Science Society of America Journal}  \bvol{77}~(5),
  \pg{1461--1477}.

\bibitem[Godr{\`{e}}che \& Luck(2001)]{Godreche2001}
{\sc \au{Godr{\`{e}}che, C.} \& \au{Luck, J.~M.}} \yr{2001}  \at{Statistics of
  the occupation time of renewal processes}.  \jt{Journal of Statistical
  Physics}  \bvol{104}~(3/4),  \pg{489--524}.

\bibitem[Hendry {\em et~al.\/}(1999)Hendry, Lawrence \&
  Maloszewski]{Hendry1999}
{\sc \au{Hendry, M.~J.}, \au{Lawrence, J.~R.} \& \au{Maloszewski, P.}}
  \yr{1999}  \at{Effects of velocity on the transport of two bacteria through
  saturated sand}.  \jt{Ground Water}  \bvol{37}~(1),  \pg{103--112}.

\bibitem[Holzner {\em et~al.\/}(2015)Holzner, Morales, Willmann \&
  Dentz]{holzner2015intermittent}
{\sc \au{Holzner, Markus}, \au{Morales, Ver{\'o}nica~L}, \au{Willmann,
  Matthias} \& \au{Dentz, Marco}} \yr{2015}  \at{Intermittent {L}agrangian
  velocities and accelerations in three-dimensional porous medium flow}.
  \jt{Physical Review E}  \bvol{92}~(1),  \pg{013015}.

\bibitem[Hornberger {\em et~al.\/}(1992)Hornberger, Mills \&
  Herman]{Hornberg1992}
{\sc \au{Hornberger, G.M.}, \au{Mills, A.L.} \& \au{Herman, J.S.}} \yr{1992}
  \at{Bacterial transport in porous media evaluation of a model using
  laboratory observation}.  \jt{Water Resources Research}  \pg{pp. 915--938}.

\bibitem[Hyman {\em et~al.\/}(2019)Hyman, Rajaram, Srinivasan, Makedonska,
  Karra, Viswanathan \& Srinivasan]{Hyman2019}
{\sc \au{Hyman, Jeffrey~D.}, \au{Rajaram, Harihar}, \au{Srinivasan, Shriram},
  \au{Makedonska, Nataliia}, \au{Karra, Satish}, \au{Viswanathan, Hari} \&
  \au{Srinivasan, Gowri}} \yr{2019}  \at{Matrix diffusion in fractured media:
  New insights into power law scaling of breakthrough curves}.  \jt{Geophysical
  Research Letters}  \bvol{46}~(23),  \pg{13785--13795}.

\bibitem[Jeffery(1922)]{Jeffery1922}
{\sc \au{Jeffery, G.B.}} \yr{1922}  \at{The motion of ellipsoidal particles
  immersed in a viscous fluid}.  \jt{Proceedings of the Royal Society of
  London. Series A, Containing Papers of a Mathematical and Physical Character}
   \bvol{102}~(715),  \pg{161--179}.

\bibitem[Jiang {\em et~al.\/}(2005)Jiang, Noonan, Buchan \&
  Smith]{Guangming2005}
{\sc \au{Jiang, Guangming}, \au{Noonan, Mike~J.}, \au{Buchan, Graeme~D.} \&
  \au{Smith, Neil}} \yr{2005}  \at{Transport and deposition of bacillus
  subtilis through an intact soil column}.  \jt{Australian Journal of Soil
  Research ,}  \bvol{43},  \pg{695--703}.

\bibitem[Jing {\em et~al.\/}(2020)Jing, Z{\"o}ttl, Cl{\'e}ment \&
  Lindner]{Jing2020}
{\sc \au{Jing, Guangyin}, \au{Z{\"o}ttl, Andreas}, \au{Cl{\'e}ment, {\'E}ric}
  \& \au{Lindner, Anke}} \yr{2020}  \at{Chirality-induced bacterial rheotaxis
  in bulk shear flows}.  \jt{Science Advances}  \bvol{6}~(28),  \pg{eabb2012}.

\bibitem[de~Josselin~de Jong(1958)]{deJong1958}
{\sc \au{de~Josselin~de Jong, G.}} \yr{1958}  \at{Longitudinal and transverse
  diffusion in granular deposits}.  \jt{Trans. Amer. Geophys. Un.}  \bvol{39},
  \pg{67--74}.

\bibitem[Junot {\em et~al.\/}(2021)Junot, Darnige, Lindner, Martinez, Arlt,
  Dawson, Poon, Auradou \& Cl{\'e}ment]{Junot2022}
{\sc \au{Junot, Gaspard}, \au{Darnige, Thierry}, \au{Lindner, Anke},
  \au{Martinez, Vincent~A.}, \au{Arlt, Jochen}, \au{Dawson, Angela}, \au{Poon,
  Wilson C.~K.}, \au{Auradou, Harold} \& \au{Cl{\'e}ment, Eric}} \yr{2021}
  Run-to-tumble variability controls the surface residence times of ${\it
  e.~coli}$ bacteria.

\bibitem[Kaya \& Koser(2012)]{Kaya2012}
{\sc \au{Kaya, Tolga} \& \au{Koser, Hur}} \yr{2012}  \at{Direct upstream
  motility in escherichia coli}.  \jt{Biophysical Journal}  \bvol{102}~(7),
  \pg{1514--1523}.

\bibitem[Koponen {\em et~al.\/}(1996)Koponen, Kataja \& Timonen]{kop1996}
{\sc \au{Koponen, A.}, \au{Kataja, M.} \& \au{Timonen, J.}} \yr{1996}
  \at{Tortuous flow in porous media}.  \jt{Phys. Rev. E}  \bvol{54},
  \pg{406--410}.

\bibitem[Lee {\em et~al.\/}(2021)Lee, Lohrmann, Szuttor, Auradou \&
  Holm]{Lee2021}
{\sc \au{Lee, Miru}, \au{Lohrmann, Christoph}, \au{Szuttor, Kai}, \au{Auradou,
  Harold} \& \au{Holm, Christian}} \yr{2021}  \at{The influence of motility on
  bacterial accumulation in a microporous channel}.  \jt{Soft Matter}
  \bvol{17}~(4),  \pg{893--902}.

\bibitem[Li {\em et~al.\/}(2011)Li, Bensson, Nisimova, Munger, Mahautmr, Tang,
  Maxey \& Brun]{Li2011}
{\sc \au{Li, Guanglai}, \au{Bensson, James}, \au{Nisimova, Liana}, \au{Munger,
  Daniel}, \au{Mahautmr, Panrapee}, \au{Tang, Jay~X.}, \au{Maxey, Martin~R.} \&
  \au{Brun, Yves~V.}} \yr{2011}  \at{Accumulation of swimming bacteria near a
  solid surface}.  \jt{Physical Review E}  \bvol{84}~(4),  \pg{041932}.

\bibitem[Liang {\em et~al.\/}(2018)Liang, Lu, Chang, Nguyen \&
  Massoudieh]{Massoudieh2018}
{\sc \au{Liang, Xiaomeng}, \au{Lu, Nanxi}, \au{Chang, Lin-Ching}, \au{Nguyen,
  Thanh~H.} \& \au{Massoudieh, Arash}} \yr{2018}  \at{Evaluation of bacterial
  run and tumble motility parameters through trajectory analysis}  \bvol{211},
  \pg{26--38}.

\bibitem[Liu {\em et~al.\/}(2011)Liu, Ford \& Smith]{Liu2011}
{\sc \au{Liu, Jun}, \au{Ford, Roseanne~M.} \& \au{Smith, James~A.}} \yr{2011}
  \at{Idling time of motile bacteria contributes to retardation and dispersion
  in sand porous medium}  \bvol{45}~(9),  \pg{3945--3951}.

\bibitem[Marcos {\em et~al.\/}(2012)Marcos, Fu, Powers \& Stocker]{Marcos2012}
{\sc \au{Marcos}, \au{Fu, Henry~C.}, \au{Powers, Thomas~R.} \& \au{Stocker,
  Roman}} \yr{2012}  \at{Bacterial rheotaxis}.  \jt{Proceedings of the National
  Academy of Sciences}  \bvol{109}~(13),  \pg{4780--4785}.

\bibitem[Margolin {\em et~al.\/}(2003)Margolin, Dentz \&
  Berkowitz]{Margolin2003}
{\sc \au{Margolin, Gennady}, \au{Dentz, Marco} \& \au{Berkowitz, Brian}}
  \yr{2003}  \at{Continuous time random walk and multirate mass transfer
  modeling of sorption}.  \jt{Chemical Physics}  \bvol{295}~(1),  \pg{71--80}.

\bibitem[Mathijssen {\em et~al.\/}(2019)Mathijssen, Figueroa-Morales, Junot,
  Cl{\'e}ment, Lindner \& Z{\"o}ttl]{Mathijssen2019}
{\sc \au{Mathijssen, Arnold J. T.~M.}, \au{Figueroa-Morales, Nuris}, \au{Junot,
  Gaspard}, \au{Cl{\'e}ment, {\'E}ric}, \au{Lindner, Anke} \& \au{Z{\"o}ttl,
  Andreas}} \yr{2019}  \at{Oscillatory surface rheotaxis of swimming e. coli
  bacteria}.  \jt{Nature Communications}  \bvol{10}~(1),  \pg{1--12}.

\bibitem[Matyka {\em et~al.\/}(2016)Matyka, Golembiewski \& Koza]{Matyka2016}
{\sc \au{Matyka, M.}, \au{Golembiewski, J.} \& \au{Koza, Z.}} \yr{2016}
  \at{Power-exponential velocity distributions in disordered porous media}.
  \jt{Phys. Rev. E}  \bvol{93},  \pg{013110}.

\bibitem[{McCaulou} {\em et~al.\/}(1994){McCaulou}, Bales \&
  {McCarthy}]{McCaulou1994}
{\sc \au{{McCaulou}, Douglas~R.}, \au{Bales, Roger~C.} \& \au{{McCarthy},
  John~F.}} \yr{1994}  \at{Use of short-pulse experiments to study bacteria
  transport through porous media}.  \jt{Journal of contaminant hydrology}
  \bvol{15}~(1),  \pg{1--14}.

\bibitem[Mi{\~n}o {\em et~al.\/}(2018)Mi{\~n}o, Baabour, Chertcoff, Gutkind,
  Cl{\'e}ment, Auradou \& Ippolito]{Mino2018}
{\sc \au{Mi{\~n}o, Gast{\'o}n~L.}, \au{Baabour, Magali}, \au{Chertcoff,
  Ricardo}, \au{Gutkind, Gabriel}, \au{Cl{\'e}ment, Eric}, \au{Auradou, Harold}
  \& \au{Ippolito, Irene}} \yr{2018}  \at{\textit{E coli} accumulation behind
  an obstacle}.  \jt{Advances in Microbiology}  \bvol{08}~(6),  \pg{451}.

\bibitem[Morales {\em et~al.\/}(2017)Morales, Dentz, Willmann \&
  Holzner]{morales2017}
{\sc \au{Morales, Veronica~L}, \au{Dentz, Marco}, \au{Willmann, Matthias} \&
  \au{Holzner, Markus}} \yr{2017}  \at{Stochastic dynamics of intermittent
  pore-scale particle motion in three-dimensional porous media: Experiments and
  theory}.  \jt{Geophysical Research Letters}  \bvol{44}~(18),
  \pg{9361--9371}.

\bibitem[Noetinger {\em et~al.\/}(2016)Noetinger, Roubinet, Russian, Le~Borgne,
  Delay, Dentz, de~Dreuzy \& Gouze]{noetinger2016}
{\sc \au{Noetinger, Benoit}, \au{Roubinet, Delphine}, \au{Russian, Anna},
  \au{Le~Borgne, Tanguy}, \au{Delay, Frederick}, \au{Dentz, Marco},
  \au{de~Dreuzy, Jean-Raynald} \& \au{Gouze, Philippe}} \yr{2016}  \at{Random
  walk methods for modeling hydrodynamic transport in porous and fractured
  media from pore to reservoir scale}.  \jt{Transport in Porous Media}
  \bvol{115}~(2),  \pg{345--385}.

\bibitem[Puyguiraud {\em et~al.\/}(2019{\natexlab{{\em a\/}}})Puyguiraud, Gouze
  \& Dentz]{puyguiraud2019a}
{\sc \au{Puyguiraud, Alexandre}, \au{Gouze, Philippe} \& \au{Dentz, Marco}}
  \yr{2019{\natexlab{{\em a\/}}}}  \at{Stochastic dynamics of lagrangian
  pore-scale velocities in three-dimensional porous media}.  \jt{Water
  Resources Research}  \bvol{55}~(2),  \pg{1196--1217}.

\bibitem[Puyguiraud {\em et~al.\/}(2019{\natexlab{{\em b\/}}})Puyguiraud, Gouze
  \& Dentz]{puyguiraud2019upscaling}
{\sc \au{Puyguiraud, Alexandre}, \au{Gouze, Philippe} \& \au{Dentz, Marco}}
  \yr{2019{\natexlab{{\em b\/}}}}  \at{Upscaling of anomalous pore-scale
  dispersion}.  \jt{Transport in Porous Media}  \bvol{128}~(2),  \pg{837--855}.

\bibitem[Puyguiraud {\em et~al.\/}(2021)Puyguiraud, Gouze \&
  Dentz]{puyguiraud2021pore}
{\sc \au{Puyguiraud, Alexandre}, \au{Gouze, Philippe} \& \au{Dentz, Marco}}
  \yr{2021}  \at{Pore-scale mixing and the evolution of hydrodynamic dispersion
  in porous media}.  \jt{Physical Review Letters}  \bvol{126}~(16),
  \pg{164501}.

\bibitem[Rusconi {\em et~al.\/}(2014)Rusconi, Guasto \& Stocker]{Rusconi2014}
{\sc \au{Rusconi, Roberto}, \au{Guasto, Jeffrey~S.} \& \au{Stocker, Roman}}
  \yr{2014}  \at{Bacterial transport suppressed by fluid shear}.  \jt{Nature
  Physics}  \bvol{10}~(3),  \pg{212--217}.

\bibitem[Saffman(1959)]{saffman1959}
{\sc \au{Saffman, PG}} \yr{1959}  \at{A theory of dispersion in a porous
  medium}.  \jt{Journal of Fluid Mechanics}  \bvol{6}~(03),  \pg{321--349}.

\bibitem[Scheidweiler {\em et~al.\/}(2020)Scheidweiler, Miele, Peter, Battin \&
  de~Anna]{Scheidweiler2020}
{\sc \au{Scheidweiler, David}, \au{Miele, Filippo}, \au{Peter, Hannes},
  \au{Battin, Tom~J.} \& \au{de~Anna, Pietro}} \yr{2020}  \at{Trait-specific
  dispersal of bacteria in heterogeneous porous environments: from pore to
  porous medium scale}.  \jt{Journal of The Royal Society Interface}
  \bvol{17}~(164),  \pg{20200046}.

\bibitem[Secchi {\em et~al.\/}(2020)Secchi, Vitale, Mi{\~{n}}o, Kantsler,
  Eberl, Rusconi \& Stocker]{Secchi2020}
{\sc \au{Secchi, Eleonora}, \au{Vitale, Alessandra}, \au{Mi{\~{n}}o,
  Gast{\'{o}}n~L.}, \au{Kantsler, Vasily}, \au{Eberl, Leo}, \au{Rusconi,
  Roberto} \& \au{Stocker, Roman}} \yr{2020}  \at{The effect of flow on
  swimming bacteria controls the initial colonization of curved surfaces}.
  \jt{Nature Communications}  \bvol{11}~(1).

\bibitem[Siena {\em et~al.\/}(2014)Siena, Riva, Hyman, Winter \&
  Guadagnini]{siena2014relationship}
{\sc \au{Siena, M}, \au{Riva, M}, \au{Hyman, JD}, \au{Winter, C~Larrabee} \&
  \au{Guadagnini, A}} \yr{2014}  \at{Relationship between pore size and
  velocity probability distributions in stochastically generated porous media}.
   \jt{Physical Review E}  \bvol{89}~(1),  \pg{013018}.

\bibitem[Souzy {\em et~al.\/}(2020)Souzy, Lhuissier, M{\'{e}}heust, Borgne \&
  Metzger]{souzy2020}
{\sc \au{Souzy, M.}, \au{Lhuissier, H.}, \au{M{\'{e}}heust, Y.}, \au{Borgne,
  T.~Le} \& \au{Metzger, B.}} \yr{2020}  \at{Velocity distributions, dispersion
  and stretching in three-dimensional porous media}.  \jt{Journal of Fluid
  Mechanics}  \bvol{891}.

\bibitem[Stumpp {\em et~al.\/}(2011)Stumpp, Lawrence, Hendry \&
  Maloszewski]{Stumpp2011}
{\sc \au{Stumpp, Christine}, \au{Lawrence, John~R.}, \au{Hendry, M.~Jim} \&
  \au{Maloszewski, Piotr}} \yr{2011}  \at{Transport and bacterial interactions
  of three bacterial strains in saturated column experiments}.
  \jt{Environmental Science \& Technology}  \bvol{45}~(6),  \pg{2116--2123}.

\bibitem[Taylor(1953)]{Taylor:1953}
{\sc \au{Taylor, G.~I.}} \yr{1953}  \at{Dispersion of soluble matter in solvent
  flowing slowly through a tube}.  \jt{Proc. R. Soc. Lond. A}  \bvol{219},
  \pg{186--203}.

\bibitem[Tufenkji(2007)]{Tufenkji2007}
{\sc \au{Tufenkji, Nathalie}} \yr{2007}  \at{Modeling microbial transport in
  porous media: Traditional approaches and recent developments}.  \jt{Advances
  in Water Resources}  \bvol{30}~(6),  \pg{1455--1469}.

\bibitem[Walker {\em et~al.\/}(2005)Walker, Redman \& Elimelech]{Walker2005}
{\sc \au{Walker, Sharon~L.}, \au{Redman, Jeremy~A.} \& \au{Elimelech,
  Menachem}} \yr{2005}  \at{Influence of growth phase on bacterial deposition:
  Interaction mechanisms in packed-bed column and radial stagnation point flow
  systems}.  \jt{Environmental Science \& Technology}  \bvol{39}~(17),
  \pg{6405--6411}.

\bibitem[Yates {\em et~al.\/}(1988)Yates, Yates \& Gerba]{Yates1988}
{\sc \au{Yates, Marylynn~V.}, \au{Yates, Scott~R.} \& \au{Gerba, Charles~P.}}
  \yr{1988}  \at{Modeling microbial fate in the subsurface environment}.
  \jt{Critical Reviews in Environmental Control}  \bvol{17}~(4),
  \pg{307--344}.

\bibitem[Zhang {\em et~al.\/}(2021)Zhang, He, Jin, Bai, Tong \& Ni]{Zhang2021}
{\sc \au{Zhang, Mengya}, \au{He, Lei}, \au{Jin, Xin}, \au{Bai, Fan}, \au{Tong,
  Meiping} \& \au{Ni, Jinren}} \yr{2021}  \at{Flagella and their properties
  affect the transport and deposition behaviors of escherichia coli in quartz
  sand}.  \jt{Environmental Science {\&} Technology}  \bvol{55}~(8),
  \pg{4964--4973}.

\end{thebibliography}
%%%%%%%%%%%%%%%%%%%%%%%%%%%%%%%%%%%%%%%%%%%%

\end{document}